\documentclass[aps,pre,preprint,superscriptaddress,showpacs]{revtex4-1}
\usepackage{amsmath}
\usepackage{amssymb}
\usepackage{graphicx}

\begin{document}

\preprint{}

\title{A scale-invariant probabilistic model based on Leibniz-like pyramids}

\author{A. Rodr\'{\i}guez}
\affiliation{GISC, Dpto. de Matem\'{a}tica Aplicada y Estad\'{\i}stica,
        Universidad Polit\'{e}cnica de Madrid, Pza. Cardenal Cisneros s/n, 28040 Madrid, Spain}
\author{C. Tsallis}
\affiliation{Centro Brasileiro de Pesquisas F\'{\i}sicas and National Institute of Science and Technology for Complex Systems, Rua Xavier Sigaud 150, 22290-180 Rio de Janeiro, Brazil}
\affiliation{Santa Fe Institute, 1399 Hyde Park Road, Santa Fe, New Mexico 87501, USA}

\date{\today}

\begin{abstract}

We introduce a family of probabilistic {\it scale-invariant} Leibniz-like pyramids and $(d+1)$-dimensional hyperpyramids ($d=1,2,3,...$), with $d=1$ corresponding to triangles, $d=2$ to (tetrahedral) pyramids, and so on. For all values of $d$, they are characterized by a parameter $\nu>0$, whose value determines the degree of correlation between $N$ $(d+1)$-valued random variables ($d=1$ corresponds to binary variables, $d=2$ to ternary variables, and so on). There are $(d+1)^N$ different events, and the limit $\nu\to\infty$ corresponds to independent random variables, in which case each event has a probability $1/(d+1)^N$ to occur. The sums of these $N$  $\,(d+1)$-valued random variables correspond to a $d-$dimensional probabilistic model, and  generalizes a recently proposed one-dimensional ($d=1$) model having $q-$Gaussians (with $q=(\nu-2)/(\nu-1)$ for $\nu \in [1,\infty)$) as $N\to\infty$ limit probability distributions for the sum of the $N$ binary variables [A. Rodr\'{\i}guez {\em et al}, J. Stat. Mech. (2008) P09006; R. Hanel {\em et al}, Eur. Phys. J. B {\bf 72}, 263 (2009)]. In the $\nu\to\infty$ limit the $d-$dimensional multinomial distribution  
is recovered for the sums, which approach a $d-$dimensional Gaussian distribution for $N\to\infty$. For any $\nu$, the conditional distributions of the $d-$dimensional model are shown to yield the corresponding joint distribution of the $(d-1)$-dimensional model with the same $\nu$.  For the $d=2$ case, we study the joint probability distribution, and identify two classes of marginal distributions, one of them being asymmetric and scale-invariant, while the other one is symmetric and only asymptotically scale-invariant. The present probabilistic model is proposed as a testing ground for a deeper understanding of the necessary and sufficient conditions for having $q$-Gaussian attractors in the $N\to\infty$ limit, the ultimate goal being a neat mathematical view of the causes clarifying the ubiquitous emergence of $q$-statistics verified in many natural, artificial and social systems.     
\end{abstract}


\maketitle

\section{Introduction} 
\label{introduccion}

In a probabilistic context, {\em scale invariance} is said to occur when for a set of $N$ random
variables, $\xi_1,\xi_2,\dots,\xi_N$, with {\em joint} probability distribution $p_N(\xi_1,\xi_2,\dots,\xi_N)$, the functional form of the {\it marginal} probabilities of a $(N-1)$-variables subset
coincides with its joint $(N-1)$-variables probability distribution, i.e, when
\begin{equation}\int
p_N(\xi_1,\xi_2,\dots,\xi_{N-1},\xi_N)\,d\xi_N=p_{N-1}(\xi_1,\xi_2,\dots,\xi_{N-1}).\label{scale_invariance}\end{equation}

In the absence of independence, this relation ---which is trivially fulfilled in the case of independent random variables--- involves the presence of {\em global} correlations, which is precisely the scenario where {\em nonextensive statistical mechanics} \cite{Tsallis:88,next:04,next:05,libro,biblio} comes to play an important role.

This theory, also referred to as $q-$statistics, generalizes the standard Boltzmann-Gibbs statistical mechanics, which in turn is appropriate to describe systems which typically present {\em local} correlations, if any. In this case, the standard Central Limit Theorem (CLT) ensures the appearance of Gaussians as attractors in the thermodynamic limit for the sums of independent or weakly correlated random variables with finite variance.

Within the framework of $q-$statistics, an extension of the CLT, the so called $q-$generalized Central Limit Theorem ($q-$CLT),  has been recently proved \cite{Umarov,Umarov2} for the case of $q-${\em independence} ---a specific class of global correlations--- which states that in this case the attractor distributions in the thermodynamic limit are the so called $q-${\em Gaussians}, which in $d$ dimensions have the form \cite{Vignat,proceedings}  
\begin{equation}G_q(\vec x)=C_{q,d}^{}\;e_q^{-\beta\; {\vec x}^T\Sigma\vec x};\quad \vec x\in \Omega_q
\label{q-gaussiana_d}\end{equation}
where $\vec x=(x_1,\dots,x_d)$,  $q$ is a real parameter with $q<1+\frac{2}{d}$, $\beta$ is a positive constant, $\Sigma$ is a positive definite matrix, $e_q^z\equiv[1+(1-q)z]^{1/(1-q)}$ $(e_1^z=e^z)$, is the so called, $q-${\em exponential} function,  whose support is $\Omega_q=\mathbb R^d$ for $q\ge1$ while $\Omega_q=\{\vec x\;/\;\vec x^T\Sigma \vec x<\frac{1}{\beta(1-q)}\}$ for $q<1$, and $C_{q,d}^{-1}=\int_{\Omega_q} e_q^{-\beta\vec x^T\Sigma \vec x}dx_1\cdots dx_d$, is the normalization constant. The $m-$th order moments of distribution \eqref{q-gaussiana_d} are defined for all $m$ if $q<1$, and only if $q<1+\frac{2}{m+d}$ for $q>1$, with $E[\vec X]=\vec 0$ and a covariance matrix given by 
\begin{equation}E[\vec X{\vec X}^{\,T}]=\frac{1}{\beta(d+4-(d+2)q)}\Sigma^{-1}\label{covarianzas_q-gaussiana_d}\end{equation}

The Gaussians (as well as the independence and the standard CLT) are recovered from Eq.~\eqref{q-gaussiana_d} for $q=1$. Another particular instance of $q-$Gaussian is the uniform distribution, which emerges in the $q\to-\infty$ limit.  

A number of recent works address the possible relationship between scale invariant correlations and  $q-$Gaussians attractors \cite{Moyano,Thistleton, Hilhorst, RodriguezSchwammleTsallis,HanelThurnerTsallis}. Though some discrete \cite{Moyano} or continuous \cite{Thistleton} scale-invariant systems have been shown {\em not} to have $q-$Gaussians, but remarkably close functions instead, as limiting probability distributions \cite{Hilhorst}, the one-dimensional scale-invariant model introduced in \cite{RodriguezSchwammleTsallis} has been analiticaly shown to yield $q-$Gaussians in the thermodynamic limit. Our goal in the present paper is to study a natural generalization of the aforementioned model to higher dimensions. 

In Sec. \ref{Scale_invariance} we go through the detailed description of scale invariance. In Sec.~\ref{one_dimensional} we briefly review the one-dimensional model based on Leibniz-like triangles. In Sec. \ref{two-dimensional} we introduce Leibniz-like pyramids and deal with the two-dimensional model. We then explore the conditional, marginal and joint probability distributions of the two-dimensional model in Sec. \ref{joint-conditional-marginal}. We generalize the model to arbitrary dimension in Sec. \ref{dimension_arbitraria}.   Finally, we summarize our conclusions in Sec. \ref{conclusions}. Some lengthy calculations are developed in the Appendix.   

\section{Scale invariance}
\label{Scale_invariance}
In order to illustrate the concept of scale invariance, let us consider a statistical model consisting of a set of $N$ identical and exchangeable binary random variables $\xi_1, \xi_2,\dots,\xi_N$, taking values $\xi^{(1)}$ and $\xi^{(2)}$. For the trivial $N=1$ case there are only $2^1=2$ events in the sample space and we have the probability distribution
\begin{equation}p_{1}(\xi_1)=r_{1,0}\delta(\xi_1-\xi^{(1)})+r_{1,1}\delta(\xi_1-\xi^{(2)})\end{equation}
where $r_{1,n}$ (with $n=0,1$, and $r_{1,0}+r_{0,1}=1$) stands for the probability that the variable $\xi_1$ take $n$ times the value $\xi^{(2)}$.

For the $N=2$ case we have $2^2=4$ different events and the corresponding probability distribution 
\begin{align}p_{2}(\xi_1,\xi_2)&=r_{2,0}\delta(\xi_1-\xi^{(1)})\delta(\xi_2-\xi^{(1)})+
r_{2,1}\delta(\xi_1-\xi^{(1)})\delta(\xi_2-\xi^{(2)})\nonumber\\&+r_{2,1}\delta(\xi_1-\xi^{(2)})\delta(\xi_2-\xi^{(1)})+r_{2,2}\delta(\xi_1-\xi^{(2)})\delta(\xi_2-\xi^{(2)})\label{N=2}\end{align}
where now the coefficients $r_{2,n}$ (with $n=0,1,2$, and $r_{2,0}+2r_{2,1}+r_{2,2}=1$) stand for the probability that the value $\xi^{(1)}$ appears $n$ times in the pair $(\xi_1,\xi_2)$. Note that in order to have exchangeable variables the probabilities associated to events $(\xi^{(2)},\xi^{(1)})$ and $(\xi^{(1)},\xi^{(2)})$ are necessarily the same.

Let us integrate now expression \eqref{N=2} with respect to variable $\xi_2$ in order to obtain the corresponding marginal distribution 
\begin{equation}\tilde p_1(\xi_1)=\int p_2(\xi_1,\xi_2)\text{d}\xi_2=(r_{2,0}+r_{2,1})\delta(\xi_1-\xi^{(1)})+(r_{2,1}+r_{2,2})\delta(\xi_1-\xi^{(2)}) \end{equation}
In order to Eq. \eqref{scale_invariance} be fulfilled, i.e., $\tilde p_1(\xi_1)=p_1(\xi_1)$, it is necessary and sufficient that $r_{2,0}+r_{2,1}=r_{1,0}$ and $r_{2,1}+r_{2,2}=r_{1,1}$.

For the $N=3$ case, with $2^3=8$ events in the sample space, the probability distribution reads
\begin{align}p_{3}(\xi_1,\xi_2,\xi_3)&=r_{3,0}\delta(\xi_1-\xi^{(1)})\delta(\xi_2-\xi^{(1)})\delta(\xi_3-\xi^{(1)})+
r_{3,1}\delta(\xi_1-\xi^{(2)})\delta(\xi_2-\xi^{(1)})\delta(\xi_3-\xi^{(1)})\nonumber\\
&+r_{3,1}\delta(\xi_1-\xi^{(1)})\delta(\xi_2-\xi^{(2)})\delta(\xi_3-\xi^{(1)})+r_{3,1}\delta(\xi_1-\xi^{(1)})\delta(\xi_2-\xi^{(1)})\delta(\xi_3-\xi^{(2)})\nonumber\\
&+r_{3,2}\delta(\xi_1-\xi^{(1)})\delta(\xi_2-\xi^{(2)})\delta(\xi_3-\xi^{(2)})+r_{3,2}\delta(\xi_1-\xi^{(2)})\delta(\xi_2-\xi^{(1)})\delta(\xi_3-\xi^{(2)})\nonumber\\
&+r_{3,2}\delta(\xi_1-\xi^{(2)})\delta(\xi_2-\xi^{(2)})\delta(\xi_3-\xi^{(1)})+r_{3,3}\delta(\xi_1-\xi^{(2)})\delta(\xi_2-\xi^{(2)})\delta(\xi_3-\xi^{(2)})\label{N=3}\end{align}
where $r_{3,n}$ (with $n=0, 1, 2, 3$, and $r_{3,0}+3r_{3,1}+3r_{3,2}+r_{3,3}=1$) stands for the probability that value $\xi^{(2)}$ appears $n$ times in the event $(\xi_1,\xi_2,\xi_3)$. Due to the exchangeability of variables, there are only $N+1=4$ different probabilities out of the $2^N=8$ different events. Integrating now with respect to $\xi_3$ in \eqref{N=3} we get the marginal distribution
\begin{align}\tilde p_2(\xi_1,\xi_2)=\int p_3(\xi_1,\xi_2,\xi_3)\text{d}\xi_3&=(r_{3,0}+r_{3,1})\delta(\xi_1-\xi^{(1)})\delta(\xi_2-\xi^{(1)})\nonumber\\&+(r_{3,1}+r_{3,2})\delta(\xi_1-\xi^{(1)})\delta(\xi_2-\xi^{(2)})\nonumber\\&+(r_{3,1}+r_{3,2})\delta(\xi_1-\xi^{(2)})\delta(\xi_2-\xi^{(1)})\nonumber\\&+(r_{3,2}+r_{3,3})\delta(\xi_1-\xi^{(2)})\delta(\xi_2-\xi^{(2)})\label{marginal_N=3}\end{align}
Now, for marginal distribution \eqref{marginal_N=3} to coincide with joint distribution \eqref{N=2} the relations between probabilities $r_{3,0}+r_{3,1}=r_{2,0}$, $r_{3,1}+r_{3,2}=r_{2,1}$  and $r_{3,2}+r_{3,3}=r_{2,2}$  must hold. 

In the general case, we have
 
\begin{equation}p_N(\xi_1,\xi_2,\dots,\xi_N)=\sum_{n=0}^Nr_{N,n}\sum_{{\cal C}\in{ C}^N_n}\delta(\xi^{}_1-\xi^{(\cal C)}_1)\delta(\xi^{}_2-\xi^{(\cal C)}_2)\cdots\delta(\xi^{}_N-\xi^{(\cal C)}_N)\label{conjunta_general}\end{equation}
where superindex $\cal C$ runs over the $\binom{N}{n}$ elements set $C^N_n$ of $n-$combinations of the $N$ elements set ${\cal N}=\left\{1,2,\dots, N\right\}$ and $\xi^{(\cal C)}_i$ (for $i\in\cal N$) equals $\xi^{(2)}$ if subindex $i$ is selected by combination $\cal C$ and $\xi^{(1)}$ otherwise. The total number of summands in Eq.~\eqref{conjunta_general} is thus $2^N$.
 
Integrating out with respect to $\xi_N$ in \eqref{conjunta_general} one gets the marginal probability distribution
\begin{equation}\tilde p_{N-1}(\xi_1,\xi_2,\dots,\xi_{N-1})=\sum_{n=0}^{N-1}(r_{N,n}+r_{N,n+1})\sum_{{\cal C}\in{ C}^{N-1}_n}\delta(\xi^{}_1-\xi^{(\cal C)}_1)\delta(\xi^{}_2-\xi^{(\cal C)}_2)\cdots\delta(\xi^{}_{N-1}-\xi^{(\cal C)}_{N-1})\end{equation} 
which coincides with the joint probability distribution $p_{N-1}(\xi_1,\xi_2,\dots,\xi_{N-1})$ in case that
\begin{equation}r_{N,n}+r_{N,n+1}=r_{N-1,n}.\label{Leibniz_rule}\end{equation}

This relation is known as the {\em Leibniz rule}, and reflects scale-invariance of exchangeable binary random variables. We will show a family of models satisfying relation \eqref{Leibniz_rule} in Sec.~\ref{one_dimensional}.

Let us consider now a set of $N$ identical exchangeable ternary variables, taking values $\xi^{(1)}$, $\xi^{(2)}$, and $\xi^{(3)}$. For the $N=1$ case we have $3^1=3$ different events and the probability distribution  
\begin{equation}p_1(\xi_1)=r_{1,0,0}\delta(\xi_1-\xi^{(1)})+r_{1,1,0}\delta(\xi_1-\xi^{(2)})+r_{1,0,1}\delta(\xi_1-\xi^{(3)})\label{N=1_ternarias}\end{equation}
where $r_{1,n,m}$, (with $n,m=0,1$; $0\leqslant n+m\leqslant 1$, and $r_{1,0,0}+r_{1,1,0}+r_{1,0,1}=1$)
stands for the probability of obtaining $n$ times the value $\xi^{(2)}$ and $m$ times the value $\xi^{(3)}$. 

For the $N=2$ case there are $3^2=9$ different events and the joint probability distribution reads
\begin{align}p_2(\xi_1,\xi_2)&=r_{2,0,0}\delta(\xi_1-\xi^{(1)})\delta(\xi_2-\xi^{(1)})+
r_{2,1,0}\delta(\xi_1-\xi^{(1)})\delta(\xi_2-\xi^{(2)})+r_{2,1,0}\delta(\xi_1-\xi^{(2)})\delta(\xi_2-\xi^{(1)})\nonumber\\&+
r_{2,0,1}\delta(\xi_1-\xi^{(1)})\delta(\xi_2-\xi^{(3)})+r_{2,0,1}\delta(\xi_1-\xi^{(3)})\delta(\xi_2-\xi^{(1)})+r_{2,1,1}\delta(\xi_1-\xi^{(2)})\delta(\xi_2-\xi^{(3)})\nonumber\\&+r_{2,1,1}\delta(\xi_1-\xi^{(3)})\delta(\xi_2-\xi^{(2)})+r_{2,2,0}\delta(\xi_1-\xi^{(2)})\delta(\xi_2-\xi^{(2)})+r_{2,0,2}\delta(\xi_1-\xi^{(3)})\delta(\xi_2-\xi^{(3)})
\end{align}
where $r_{2,n,m}$ (with $n,m=0, 1, 2$; $0\leqslant n+m\leqslant 2$, and $r_{2,0,0}+r_{2,2,0}+r_{2,0,2}+2(r_{2,0,1}+r_{2,1,0}+r_{2,1,1})=1$) stands for the probability of obtaining $n$ times the value $\xi^{(2)}$ and $m$ times the value $\xi^{(3)}$ in event $(\xi_1,\xi_2)$. Due to exchangeability, the number of different probabilities reduces to $\frac{(N+1)(N+2)}{2}=6$. 

Integrating now with respect to $\xi_2$ one gets the marginal distribution
\begin{align}\tilde p_1(\xi_1)=\int p_2(\xi_1,\xi_2)\text{d}\xi_2&=(r_{2,0,0}+r_{2,0,1}+r_{2,1,0})\delta(\xi_1-\xi^{(1)})\nonumber\\&+
(r_{2,1,0}+r_{2,1,1}+r_{2,1,1})\delta(\xi_1-\xi^{(2)})\nonumber\\&+
(r_{2,0,1}+r_{2,0,2}+r_{2,1,1})\delta(\xi_1-\xi^{(3)})\end{align} 
which coincides with the $N=1$ distribution \eqref{N=1_ternarias} when $r_{2,0,0}+r_{2,0,1}+r_{2,1,0}=r_{1,0,0}$, $r_{2,1,0}+r_{2,1,1}+r_{2,1,1}=r_{1,1,0}$ and $r_{2,0,1}+r_{2,0,2}+r_{2,1,1}=r_{1,0,1}$.  

In the general case, we have
\begin{equation}p_N(\xi_1,\xi_2,\dots,\xi_N)=\sum_{n+m=0}^Nr_{N,n,m}\sum_{{\cal C}\in{ C}^N_{n,m}}\delta(\xi^{}_1-\xi^{(\cal C)}_1)\delta(\xi^{}_2-\xi^{(\cal C)}_2)\cdots\delta(\xi^{}_N-\xi^{(\cal C)}_N)\label{conjunta_general_ternarias}\end{equation}
where superindex $\cal C$ runs over the $\binom{N}{n,m}$ elements set $C^N_{n,m}$ of $(n,m)-$combinations of the $N$ elements set ${\cal N}=\left\{1,2,\dots, N\right\}$, and $\xi^{(\cal C)}_i$ (for $i\in\cal N$) equals $\xi^{(2)}$ if subindex $i$ belongs to the $n$ elements subset of combination $\cal C$, $\xi^{(3)}$ if subindex $i$ belongs to the $m$ elements subset of combination $\cal C$, and $\xi^{(1)}$ otherwise.
After integrating with respect to $\xi_N$ in \eqref{conjunta_general_ternarias} the following marginal probability distribution is obtained
\begin{align}\tilde p_{N-1}(\xi_1,\xi_2,\dots,\xi_{N-1})&=\sum_{n+m=0}^{N-1}(r_{N,n,m}+r_{N,n+1,m-1}+r_{N,m,m-1})
\nonumber\\&\times\sum_{{\cal C}\in{ C}^{N-1}_{n,m}}\delta(\xi^{}_1-\xi^{(\cal C)}_1)\delta(\xi^{}_2-\xi^{(\cal C)}_2)\cdots\delta(\xi^{}_{N-1}-\xi^{(\cal C)}_{N-1}).\end{align} 
This distribution coincides with the $N-1$ variables joint probability distribution if
\begin{equation}r_{N,n,m}+r_{N,n+1,m-1}+r_{N,n,m-1}=r_{N-1,n,m-1}
\label{generalized_Leibniz_rule}\end{equation} 
We shall refer to condition \eqref{generalized_Leibniz_rule} as the {\em generalized Leibniz rule}. We shall show a family of models satisfaying such relation in Sec.~\ref{two-dimensional}.

In the case of $N$ identical interchangeable $(d+1)-${\em ary} variables taking values $\xi^{(1)},\xi^{(2)},\dots,\xi^{(d+1)}$, the joint probability distribution reads
\begin{equation}p_N(\xi_1,\xi_2,\dots,\xi_N)=\sum_{n_1+\cdots+ n_d=0}^Nr_{N,n_1,n_2,\dots,n_d}\sum_{{\cal C}\in{ C}^N_{n_1,n_2,\dots,n_d}}\delta(\xi^{}_1-\xi^{(\cal C)}_1)\delta(\xi^{}_2-\xi^{(\cal C)}_2)\cdots\delta(\xi^{}_N-\xi^{(\cal C)}_N)\label{conjunta_general_general}\end{equation}
where superindex $\cal C$ runs over the $\binom{N}{n_1,\dots,n_d}$ elements set $C^N_{n_1,n_2,\dots,n_d}$ of $(n_1,n_2,\dots,n_d)-$combinations of the $N$ elements set ${\cal N}=\left\{1,2,\dots, N\right\}$ and $\xi^{(\cal C)}_i$ (for $i\in\cal N$) equals $\xi^{(j+1)}$ if subindex $i$ belongs to the $n_j$ elements subset corresponding to $\cal C$, for $j=1,\dots,d$, and $\xi^{(1)}$ otherwise.
For $d\geqslant 2$, the joint probability distribution \eqref{conjunta_general_general} satisfy scale invariance condition \eqref{scale_invariance} if ($\vec n\equiv(n_1,n_2,\dots,n_d)$):
\begin{equation}r^{(\nu)}_{N,\vec n}+r^{(\nu)}_{N,\vec n+\vec\varepsilon_1}+\cdots+r^{(\nu)}_{N,\vec n+\vec\varepsilon_d}=r^{(\nu)}_{N-1,\vec n+\vec\varepsilon_{d}}\label{scale_invariance_general}\end{equation}
where $\vec \varepsilon_1=\vec e_1-\vec e_2$, $\vec\varepsilon_i=\vec e_{i+1}-\vec e_2$, for $i=2,\dots,d-1$, and $\vec\varepsilon_d=-\vec e_2$; $\vec e_i=(0,\dots,1,\dots,0)$ being a $d-$dimensional vector whose only nonzero component is the $i-$th one, taking value 1. Relation \eqref{scale_invariance_general} reflects the scale-invariance of exchangeable $(d+1)-$nary random variables.

We shall dedicate the following sections to the detailed description of scale invariant probabilistic models. 
 
\section{One-dimensional model revisited}
\label{one_dimensional}

Let us consider a random experiment which consists in flipping $N$ biased coins. We shall call $X_{(1)}$ the random variable that counts the number of, say, heads. $X_{(1)}\in\{0,1,\dots,N\}$ can be written as 
\begin{equation}X_{(1)}=X_1+X_2+\dots+X_N\label{X}\end{equation} 
where each $X_i$, $i=1,\dots,N$, is a binary random variable taking values 1 (head) with probability $p$, and 0 (tail) with probability $(1-p)$. As it is well known, considering that different throwings are independent of each other, we have $\sigma^2_{X_iX_j}=p(1-p)\delta_{ij}$, and the probability of obtaining $n$ heads in $N$ trials is given by
\begin{equation}P(X_{(1)}=n)\equiv p_{N,n}=\binom{N}{n}p^n(1-p)^{N-n}\label{binomial}\end{equation}    
with $\sum_{n=0}^Np_{N,n}=1$, which is no other than the binomial distribution, i.e.,  $X_{(1)}\sim B(N,p)$, with $\langle X_{(1)}\rangle=Np$ and $\sigma^2_{X_{(1)}}=Np(1-p)$, where the binomial coefficients $\binom{N}{n}$ stand for the different ways in which the $n$ heads can be obtained. In other words, though there are $\Omega(N)=2^N$ different events in the sample space, only $N+1$ among them are assigned different probability values, namely $r_{N,n}\equiv p^n(1-p)^{N-n}$, $n=0,1,\dots,N$, (it is irrelevant which specific coins yield head, $X_{(1)}$ only counts the number of them). These selected probability values may be displayed in a triangle in the form 
\begin{center}\begin{tabular}{ccccccccccccc}
&&&&&&&$r_{N,n}$&&&&\\\\
$(N=0)\;$&&&&&&&1&&&&\\
$(N=1)\;$&&&&&&$1-p$&&$p$&&&&\\
$(N=2)\;$&&&&&$(1-p)^2$&&$(1-p)p$&&$p^2$&&&\\
$(N=3)\;$&&&&$(1-p)^3$&&$(1-p)^2p$&&$(1-p)p^2$&&$\;\quad p^3$&&\\
&&&&&&$\vdots$&&$\vdots$&&&&
\end{tabular}\end{center}
where the $N$-th row displays the $N+1$ probabilities, $r_{N,n}$, for $n=0,1,\dots,N$. In order to get the actual probabilities \eqref{binomial} the above triangle has to be multiplied, element by element,  by the Pascal triangle.

Let us now show explicitly the scale invariant character (which otherwise follows trivially due to the independent character of the variables) of our probabilistic model. Any of the binary variables follows a Bernoulli distribution, $p_1(X_i=x_i)=p^{x_i}(1-p)^{1-x_i}$ with $x_i=0$ or 1, and $i=1,\dots,N$. Thus, as the $N$ variables are independent, we have 
\begin{equation}p_N(x_1,x_2,\dots,x_N)=\prod_{i=1}^Np_1(x_i)=p^{\;\;\displaystyle\sum_{i=1}^Nx_i}(1-p)^{\displaystyle N-\sum_{i=1}^Nx_i}\end{equation}  
Considering now the marginal distribution corresponding to the first $(N-1)$ variables one gets
\begin{align}\sum_{x_N=0}^1p_N(x_1,x_2,\dots,x_N)&=p^{\;\;\displaystyle\sum_{i=1}^{N-1}x_i}(1-p)^{\displaystyle N-\sum_{i=1}^{N-1}x_i}+p^{\displaystyle1+\sum_{i=1}^{N-1}x_i}(1-p)^{\displaystyle N-\left(1+\sum_{i=1}^{N-1}x_i\right)}\nonumber\\&=p^{\;\;\displaystyle\sum_{i=1}^{N-1}x_i}(1-p)^{\displaystyle N-\sum_{i=1}^{N-1}x_i}\left(1+\frac{p}{1-p}\right)\label{marginal}\\&=p^{\;\;\displaystyle\sum_{i=1}^{N-1}x_i}(1-p)^{\displaystyle N-1-\sum_{i=1}^{N-1}x_i}=p_{N-1}(x_1,x_2,\dots,x_{N-1}),\nonumber\end{align}
so a discrete version of Eq. \eqref{scale_invariance} is fulfilled and the model is scale-invariant. In addition, it is readily seen that coefficients $r_{N,n}$ displayed in the above triangle follow the Leibniz rule (Eq.~\eqref{Leibniz_rule}, which, as stated in Sec.~\ref{Scale_invariance}, serves as an alternative description of scale invariance for binary variables). Thus, following Eq.~\eqref{Leibniz_rule}, the sum of two consecutive coefficients in any row of the triangle yields the coefficient on top of them.

Let us recall that the CLT states that, after  properly centering and rescaling, one obtains a Gaussian (which can be seen as a $q-$Gaussian with $q=1$) distribution out of \eqref{binomial}, namely $\frac{X_{(1)}-Np}{\sqrt{Np(1-p)}}\sim{\cal N}(0,1)$ for $N\to\infty$. We shall try now to modify the model so as to obtain $q-$Gaussians with $q\neq 1$ in the thermodynamic limit.

\subsection{Scale-invariant triangles}
\label{scale_invariant_triangles}
So far, we have considered independent variables. We shall now introduce correlations in the model in such a way that the Leibniz triangle rule \eqref{Leibniz_rule} is preserved ---that is, scale invariant correlations--- by substituting the probabilities $r_{N,n}=p^n(1-p)^{N-n}$ in \eqref{binomial} by appropriate ones. In other words, we shall change the above triangle by 
another one also satisfying \eqref{Leibniz_rule}. As a first attempt, we may resort to the so called {\em Leibniz harmonic triangle} \cite{Polya}, whose coefficients, defined us 
\begin{equation}r_{N,n}^{(1)}\equiv\frac{1}{(N+1)\binom{N}{n}}=B(N-n+1,n+1);\quad n=0,1,\dots,N
\label{Leibniz_triangle}\end{equation}  
where $B(x,y)$ is the Beta function, may be displayed in triangular form us
\begin{center}\begin{tabular}{ccccccccccccccccc}
&&&&&&&&$r_{N,n}^{(1)}$&&&&&&&\\
&&&&&&&&&&&&&&&\\
$(N=0)\quad\quad$&&&&&&&&1&&&&&&&\\
%\vspace*{-.25cm}&&&&&&&&&&&&&&\\
$(N=1)\quad\quad$&&&&&&&$\dfrac{1}{2}$&&$\dfrac{1}{2}$&&&&&&\\
$(N=2)\quad\quad$&&&&&&$\dfrac{1}{3}$&&$\dfrac{1}{6}$&&$\dfrac{1}{3}$&&&&&\\
$(N=3)\quad\quad$&&&&&$\dfrac{1}{4}$&&$\dfrac{1}{12}$&&$\dfrac{1}{12}$&&$\dfrac{1}{4}$&&&&\\
$(N=4)\quad\quad$&&&&$\dfrac{1}{5}$&&$\dfrac{1}{20}$&&$\dfrac{1}{30}$&&$\dfrac{1}{20}$&&$\dfrac{1}{5}$&&&\\
$(N=5)\quad\quad$&&&$\dfrac{1}{6}$&&$\dfrac{1}{30}$&&$\dfrac{1}{60}$&&$\dfrac{1}{60}$&&$\dfrac{1}{30}$&&$\dfrac{1}{6}$&&\\
&&&&$\vdots$&&&&$\vdots$&&&&$\vdots$&
\end{tabular}\end{center}
Making use of the properties of the Beta function it is a simple task checking that the Leibniz triangle above satisfies the rule \eqref{Leibniz_rule} which, for this reason, is named Leibniz rule. 

Substituting now the $r_{N,n}$ coefficients by the Leibniz coefficients \eqref{Leibniz_triangle} in Eq. \eqref{binomial} we obtain
\begin{equation}p^{(1)}_{N,n}\equiv\binom{N}{n}r^{(1)}_{N,n}=\frac{1}{N+1},\end{equation}
with $\sum_{n=0}^Np^{(1)}_{N,n}=1$, that is, a uniform distribution, which,  in the continuum limit, corresponds, as already mentioned in Sec. \ref{introduccion}, to a $q-$Gaussian with $q \to -\infty$.

A family of scale-invariant triangles can be now obtained as properly normalized subtriangles of the Leibniz triangle in the following fashion. Take the central coefficient of any even row of the Leibniz triangle and divide the whole triangle by it so as to turn the said coefficient to one. Now take this coefficient as the vertex of a new triangle starting downwards from it. The coefficients $r^{(\nu)}_{N,n}$ of this new triangle starting from the $2(\nu-1)-$th row of the Leibniz triangle can be then expressed as a function of the coefficients $r^{(1)}_{N,n}$ of the Leibniz triangle as
\begin{equation}r^{(\nu)}_{N,n}=\frac{r^{(1)}_{N+2(\nu-1), n+\nu-1}}{r^{(1)}_{2(\nu-1),\nu-1}}\label{Leibniz_recursivo}\end{equation}

As the only transformation we have made on the coefficients of the Leibniz triangle is a multiplication by a constant factor, Eq. \eqref{Leibniz_rule} still holds for the coefficients \eqref{Leibniz_recursivo}, thus
\begin{equation}r^{(\nu)}_{N,n}+r^{(\nu)}_{N,n+1}=r^{(\nu)}_{N-1,n}\label{Leibniz_rule_nu}\end{equation}
for any positive integer $\nu$. In virtue of Eq.~\eqref{Leibniz_rule_nu}, different coefficients of the triangle are not independent of each other and it suffices to specify one element of each row (for instance the left side of the triangle) to completely determine the whole triangle. 

Expressing now the Leibniz triangle coefficients in \eqref{Leibniz_recursivo} in terms of the Beta function as in \eqref{Leibniz_triangle}, one easily gets 
\begin{equation}r_{N,n}^{(\nu)}=\frac{B(N-n+\nu, n+\nu)}{B(\nu,\nu)};\quad\nu>0\label{triangulos_nu}\end{equation}
where now $\nu$ may take any positive value (see details in \cite{HanelThurnerTsallis}).

Apart from its symmetry, it is worth noticing another remarkable property of coefficientes \eqref{triangulos_nu}, namely, scale invariance condition \eqref{Leibniz_rule_nu} guarantees that the set of corresponding probability distributions
\begin{equation}p^{(\nu)}_{N,n}=\binom{N}{n}r^{(\nu)}_{N,n}\label{probabilidad_nu}\end{equation}
associated to the new set of variables
\begin{equation}X^{(\nu)}_{(1)}=X^{(\nu)}_1+X^{(\nu)}_2+\dots+X^{(\nu)}_N\label{X_nu}\end{equation} 
are well defined, that is, the normalization condition, $\sum_{n=0}^Np^{(\nu)}_{N,n}=1$, holds for any value of $\nu$. It may be shown that $\langle X^{(\nu)}_{(1)}\rangle=\frac{N}{2}$ for all $\nu$ and $\sigma^2_{X^{(\nu)}_{(1)}}=\frac{N(N+2\nu)}{4(1+2\nu)}$.  Concerning the binary variables in the sum \eqref{X_nu} (taking values 0 and 1 as in the independent case), it may be shown that now $\langle X^{(\nu)}_i\rangle=r^{(\nu)}_{1,0}=\frac{1}{2}\;\forall\nu$, $\sigma^2_{X^{(\nu)}_i}=\frac{1}{4}\;\forall \nu$ and $\sigma^2_{X^{(\nu)}_iX^{(\nu)}_j}=r^{(\nu)}_{2,0}-\frac{1}{4}=\frac{1}{4(1+2\nu)}$ for $i\neq j$. The constant value of the correlations, which are independent of the system size $N$ and the pair of chosen variables $X_i^{(\nu)}, X_j^{(\nu)}$, reveals the scale-invariant character of the model.

\begin{figure}
\centering\includegraphics[height=\linewidth,angle=-90,clip=]{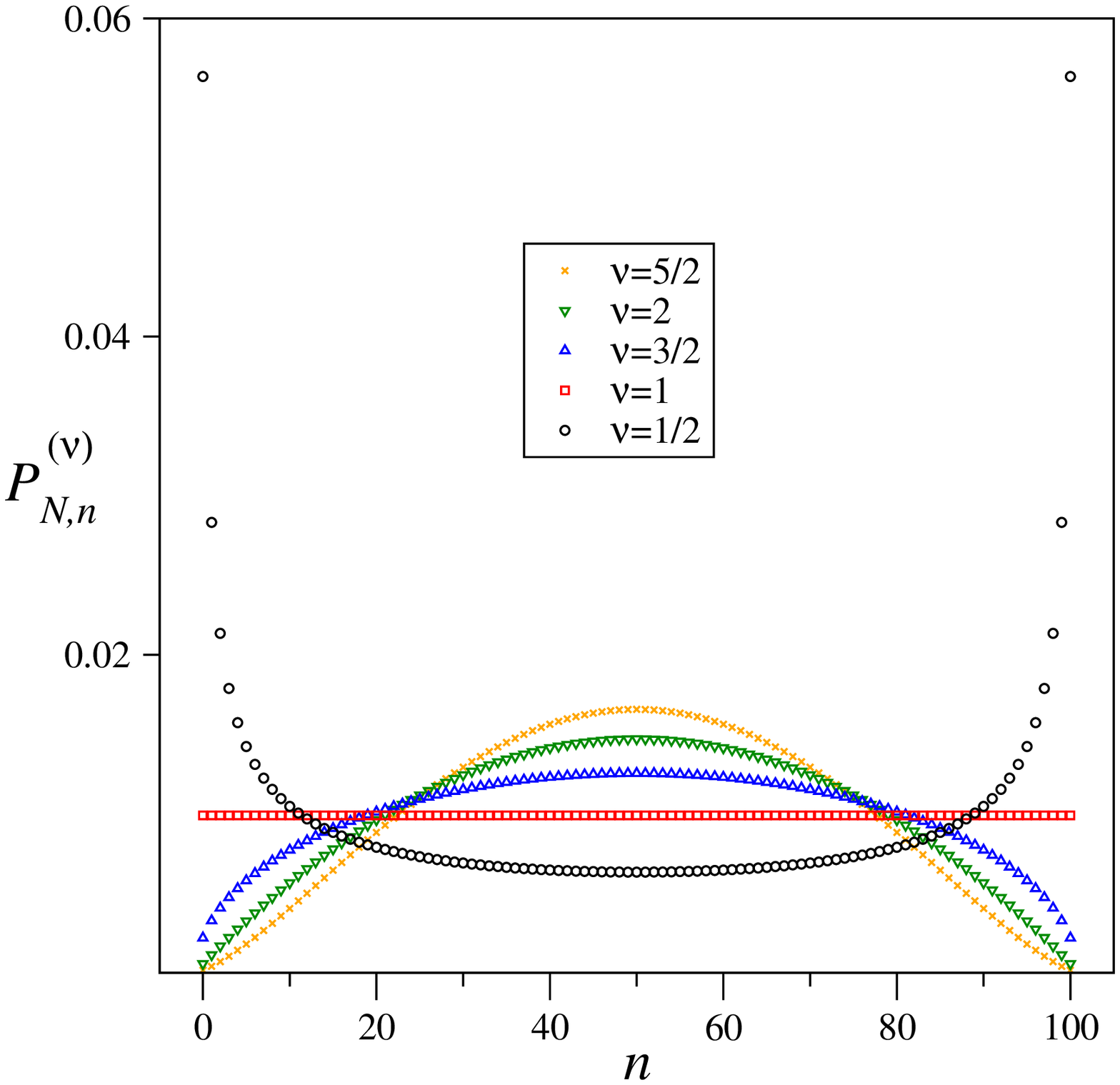}
\caption{(Color online) Probability distributions \eqref{probabilidad_nu} for $N=100$ and $\nu=\frac{1}{2}$, 1, $\frac{3}{2}$, 2 and $\frac{5}{2}$.\label{probabilidades_1d}}
\end{figure}

Probability distributions \eqref{probabilidad_nu} are shown in Fig. \ref{probabilidades_1d} for $N=100$ and $\nu=\frac{1}{2}$, 1, $\frac{3}{2}$, 2 and $\frac{5}{2}$.  

\subsection{Boltzmann and thermodynamic limits}
\label{Boltzmann_1d}
The limits $\nu\to\infty$ (that we shall call {\em Boltzmann limit} for a reason to be clear soon) and $N\to\infty$ (thermodynamic limit) are interchangeable in \eqref{triangulos_nu} (equivalently in \eqref{probabilidad_nu}). In effect, in the first case on has $\lim_{\nu\to\infty}r^{(\nu)}_{N,n}=\frac{1}{2^N}$, so 
\begin{equation}\lim_{\nu\to\infty}p^{(\nu)}_{N,n}=\binom{N}{n}\frac{1}{2^N}\label{binomial_nu_infinito}\end{equation} 
and a binomial distribution with $p=\frac{1}{2}$ (fair coin), associated to variable $X_{(1)}=\lim_{\nu\to\infty}X^{(\nu)}_{(1)}$  is recovered, which is consistent with the fact that $\lim_{\nu\to\infty}\sigma^2_{X^{(\nu)}_iX^{(\nu)}_j}=0$, thus having independent coins. Then, as the CLT states, by properly centering and rescaling variable $X_{(1)}$, one gets a Gaussian out of \eqref{binomial_nu_infinito} in the thermodynamic limit.

The limit $N\to\infty$ is much more subtle. It was studied in \cite{RodriguezSchwammleTsallis} and later extended in \cite{HanelThurnerTsallis}. Depending on the change of variable used when passing from the discrete to the continuous model, two families of $q-$Gaussians are obtained out of \eqref{probabilidad_nu} in the thermodynamic limit, with values of $q$ given by $q_\nu=\frac{\nu-2}{\nu-1}$ for $\nu>1$ or $\bar q_\nu=\frac{2\nu+3}{2\nu+1}$ for $\nu>0$. Note that two different {\em conjugated} $q-$Gaussians exist for each $\nu>1$ \cite{HanelThurnerTsallis}. In any case, one gets $\lim_{\nu\to\infty}q_\nu=\lim_{\nu\to\infty}\bar q_\nu=1$, so the ordinary Gaussian is recovered again.

\begin{figure}
\centering\includegraphics[height=\linewidth,angle=-90,clip=]{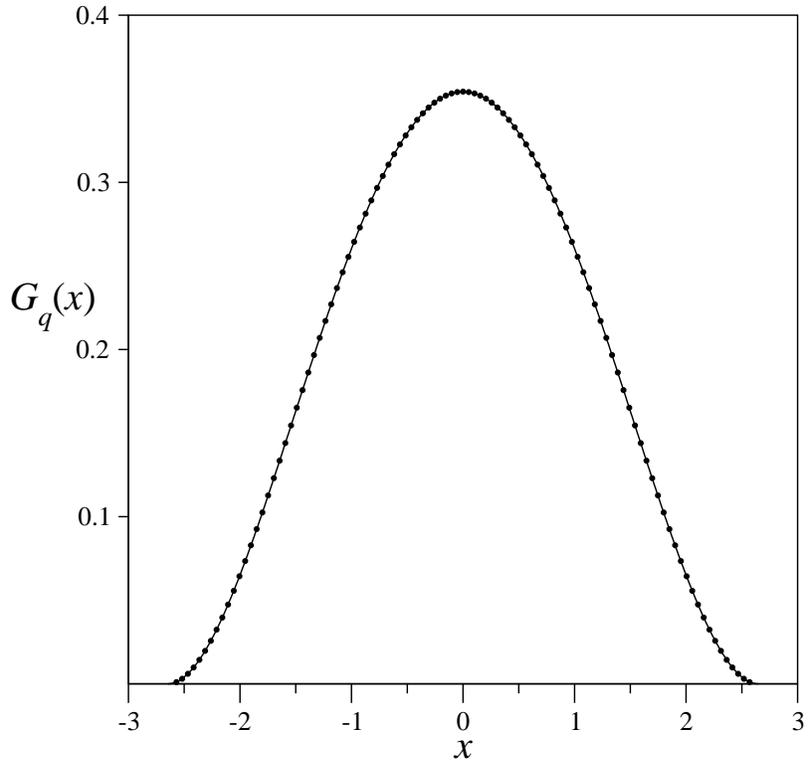}
\caption{$\sigma_\nu p^{(\nu)}_{N,n}$ versus $(n-N/2)/\sigma_\nu$ for $\nu=3$ and $N=100$ (dots) compared with the corresponding $q-$Gaussian with $q=\frac{1}{2}$ and unit variance (solid line).\label{ajuste_gaussiana}}
\end{figure}

To further support our claim, we typify variable $X_{(1)}$ by making the change 
\begin{equation}n\to u=\frac{n-\frac{N}{2}}{\sigma_\nu}\end{equation}    
with $\sigma_\nu\equiv\sqrt\frac{N(N+2\nu)}{4(1+2\nu)}$ and compare its probability distribution with the corresponding one-dimensional $q_\nu-$Gaussian \eqref{q-gaussiana_d} with $q_\nu=\frac{\nu-2}{\nu-1}$ and $\beta=\frac{1}{5-3q_\nu}$ so as to obtain unit variance
(see Eq.~\eqref{covarianzas_q-gaussiana_d}). Fig. \ref{ajuste_gaussiana} shows 
$\sigma_\nu p^{(\nu)}_{N,n}$ versus $(n-N/2)/\sigma_\nu$ for $\nu=3$ and $N=100$ (dots) compared with the corresponding $q-$Gaussian $G_q(x)$ with $q=\frac{1}{2}$ and unit variance (solid line). The agreement is surprisingly good, even for smaller (not shown) values of $N$.

\section{Two-dimensional model}
\label{two-dimensional}

Let us now generalize our random experiment to one with three different outcomes, which is equivalent to throw a biased three-sided dice. Let us label the sides $A$, $B$ and $C$ and suppose that the associated probabilities are $P(A)=p_1$, $P(B)=p_2$, with $p_1+p_2<1$, and $P(C)=1-p_1-p_2$. To properly describe our model we need to define a two components random variable 
\begin{equation}\vec X_{(2)}=\vec X_1+\vec X_2+\cdots+\vec X_N\label{vec_X}\end{equation} 
with $\vec X_{(2)}\equiv(X,Y)$, as a sum of $N$ ternary variables $\vec X_i\equiv(X_i,Y_i)$, for $i=1,\dots,N$, with values $\vec X_i=(1,0)$ (side $A$), $\vec X_i=(0,1)$ (side $B$) and $\vec X_i=(0,0)$ (side $C$). Thus $X\in\left\{0,1,\dots,N\right\}$ counts the number of $A$'s and $Y\in\left\{0,1,\dots,N\right\}$, with $0\leqslant X+Y\leqslant N$, counts the number of $B$'s out of $N$ throwings. For independent dices the $\vec X_i$ variables are independent of each other, though variables $X_i$, $Y_i$ within each pair are {\em not} ($\langle X_i\rangle=\langle X_i^2\rangle=p_1$, $\langle Y_i\rangle=\langle Y_i^2\rangle=p_2$, $\sigma^2_{X_iX_j}=p_1(1-p_1)\delta_{ij}$, $\sigma^2_{Y_iY_j}=p_2(1-p_2)\delta_{ij}$, $\sigma^2_{X_iY_j}=-p_1p_2\delta_{ij}$), neither the variables $X$ and $Y$, for which $\langle \vec X\rangle=N(p_1, p_2)$, and the covariance matrix is given by 
\begin{equation}\Sigma_{(2)}=N\left(\begin{array}{cc}p_1(1-p_1)&-p_1p_2
\\-p_1p_2&p_2(1-p_2)\end{array}\right)\label{covarianzas}\end{equation}  
The probability of having $n$ $A$'s and $m$ $B$'s, with $0\leqslant n+m\leqslant N$, in $N$ throwings is given by
\begin{equation} P(X=n,Y=m)\equiv p_{N,n,m}=\binom{N}{n,m}p_1^np_2^m(1-p_1-p_2)^{N-n-m}\label{distribucion_trinomial}
\end{equation}
with $\sum_{0\leqslant n+m\leqslant N}p_{N,n,m}=1$, which is the trinomial distribution, i.e., $\vec X_{(2)}\sim T(N,p_1,p_2)$, where the trinomial coefficients, $\binom{N}{n,\,m}=\binom{N}{n}\binom{N-n}{m}$, take into account the degeneracy since among the $\Omega(N)=3^N$ different events in the sample space only $\frac{(N+1)(N+2)}{2}$ of them have different probabilities, namely $r_{N,n,m}\equiv p_1^np_2^m(1-p_1-p_2)^{N-n-m}$. These probability values may be displayed in a pyramid

\begin{scriptsize}
\begin{center}
\begin{tabular}{lcccc}
       &&&&\begin{tabular}{c}$r_{N,n,m}$\end{tabular}\\\\
$(N=0)$&&&&\begin{tabular}{c}1\end{tabular}\\\\
$(N=1)$&&&&\begin{tabular}{ccc}&$1-p_1-p_2$&\\$p_2\quad$&&$\quad p_1$\end{tabular}\\\\
$(N=2)$&&&&\begin{tabular}{ccccc}&&$(1-p_1-p_2)^2$&&\\&$p_2(1-p_1-p_2)$&&$p_1(1-p_1-p_2)$&\\$p_2^2\quad\;$&&$p_2p_1$&&$\quad p_1^2$\end{tabular}\\\\
$(N=3)$&&&&\begin{tabular}{ccccccc}&&&$(1-p_1-p_2)^3$&&&\\&&$p_2(1-p_1-p_2)^2$&&$p_1(1-p_1-p_2)^2$&&\\
&$p_2^2(1-p_1-p_2)$&&$p_2p_1(1-p_1-p_2)$&&$p_1^2(1-p_1-p_2)$&\\$p_2^3\quad\;$&&$p_2^2p_1$&&$p_2p_1^2$&&$\quad p_1^3$\\
&$\vdots$&&$\vdots$&&$\vdots$&\end{tabular}
\end{tabular} 
\end{center}  
\end{scriptsize}
where each layer, corresponding to a different value of $N$, is a triangle
with $\frac{(N+1)(N+2)}{2}$ elements, whose rows, from top to bottom, correspond to increasing values of $n+m$ from 0 to $N$, while the ascending (descending) diagonals, from left to right (right to left), correspond to increasing values of $n$ ($m$) from 0 to $N$.  

In order to get the actual probabilities \eqref{distribucion_trinomial} the above pyramid has to be multiplied by the so called {\em Pascal pyramid}

\begin{center}
\begin{tabular}{ccccccc}
\begin{tabular}{lcccc}
       &&&&\begin{tabular}{c}$\binom{N}{n,m}$\end{tabular}\\\\
$(N=0)$&&&&\begin{tabular}{c}1\end{tabular}\\\\
$(N=1)$&&&&\begin{tabular}{ccc}&1&\\1&&1\end{tabular}\\\\
$(N=2)$&&&&\begin{tabular}{ccccc}&&1&&\\&2&&2&\\1&&2&&1\end{tabular}\\\\
$(N=3)$&&&&\begin{tabular}{ccccccc}&&&1&&&\\&&3&&3&&\\&3&&6&&3&\\1&&3&&3&&1\end{tabular}
\end{tabular}&&&&&&
\begin{tabular}{lcccc}
&&&&\\
&&&&\\
&&&&\\
$(N=4)$&&&&
\begin{tabular}{ccccccccc}
&&&&1&&&&\\
&&&4&&4&&&\\
&&6&&12&&6&&\\
&4&&12&&12&&4&\\
1&&4&&6&&4&&1\\
\end{tabular}\\\\
$(N=5)$&&&&
\begin{tabular}{ccccccccccc}
&&&&&1&&&&&\\
&&&&5&&5&&&&\\
&&&10&&20&&10&&&\\
&&10&&30&&30&&10&&\\
&5&&20&&30&&20&&5&\\
1&&5&&10&&10&&5&&1\\
&&&$\vdots$&&$\vdots$&&$\vdots$&&&

\end{tabular}
\end{tabular}
\end{tabular}
\end{center}  

\noindent which displays the trinomial coefficients $\binom{N}{n,m}$ (for the values of $n$ and $m$ within each layer the same comments as above hold).

\begin{figure}
\includegraphics[width=10cm]{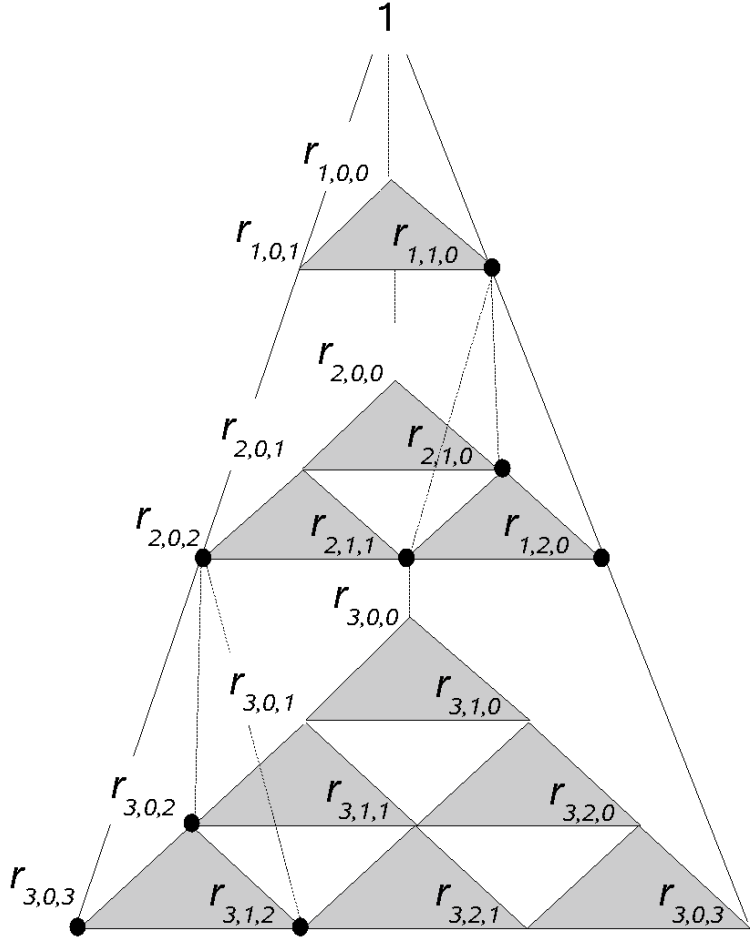}
\caption{Structure of the pyramid formed by coefficients $r_{N,n,m}=p^nq^m(1-p-q)^{N-n-m}$ (or coefficients of the family \eqref{tetraedro_beta}). Only the shaded triangles contribute to the generalized Leibniz rule. The four coefficients involved in Eq.~\eqref{generalized_Leibniz_rule} form in turn a subpyramid of the pyramid. Two of such subpyramids are indicated in de Figure.\label{figura_tetraedro}}
\end{figure}

The joint probability distribution for the independent ternary variables $\vec X_i$ is now $p_N(\vec x_1,\vec x_2,\dots,\vec x_N)=\prod_{i=1}^Np_1(\vec x_i)$, with $p_1(\vec x_i)=p^{x_i}q^{y_i}(1-p_1-p_2)^{1-x_i-y_i}$, and $\vec x_i=(1,0)$, $(0,1)$ or $(0,0)$. In an analogous fashion as in Eq.~\eqref{marginal}, it may be shown that $\sum_{\vec x_N}p_N(\vec x_1,\vec x_2,\dots,\vec x_N)=p_{N-1}(\vec x_1,\vec x_2,\dots,\vec x_{N-1})$, so the model, as mandated by the independence of the variables, still remains scale-invariant. In turn, it is easily seen that coefficientes $r_{N,n,m}$ follow the generalized Leibniz rule \eqref{generalized_Leibniz_rule}, which, as stated in Sec.~\ref{Scale_invariance},  governs scale invariance for ternary variables. 
Thus, following Eq.~\eqref{generalized_Leibniz_rule}, adding up three elements forming specific triangles within a layer (only those triangles formed by two consecutive elements of any row and the one on top of them),  
one gets the coefficient just on top of them (belonging to the above layer). Fig.~\ref{figura_tetraedro} shows the first four layers of the pyramid. Only the shaded triangles contribute to the generalized Leibniz rule. Two particular sets of four elements satisfying the generalized Leibniz rule are shown.   

In analogy with the one-dimensional case, the corresponding bidimensional version of the CLT states that the trinomial distribution yields a bidimensional Gaussian in the thermodynamic limit, that is, for $N\to\infty$ we have $\frac{\vec X_{(2)}-\langle\vec X_{(2)}\rangle}{\sqrt N}\sim{\cal N}(\vec 0,\frac{1}{N}\Sigma)$, with $\Sigma$ already given in \eqref{covarianzas}. 
Our next step is to analyze under what conditions this type of model can provide bidimensional $q-$Gaussians.

\subsection{Scale-invariant pyramids}  
As we did in Sec. \ref{scale_invariant_triangles}, we shall now introduce scale-invariant correlations in the model, so different dice throwings will no longer be independent. With this purpose, we may extend the Leibniz triangle to a {\em Leibniz-like} (tetrahedral) pyramid: 

\begin{center}
\begin{tabular}{ccccc}
\begin{tabular}{lcccc}
       &&&&\begin{tabular}{c}$r^{(1)}_{N,n,m}$\end{tabular}\\\\
$(N=0)$&&&&\begin{tabular}{c}1\end{tabular}\\\\
$(N=1)$&&&&\begin{tabular}{ccc}&$\frac{1}{3}$&\\$\frac{1}{3}$&&$\frac{1}{3}$\end{tabular}\\\\
$(N=2)$&&&&\begin{tabular}{ccccc}&&$\frac{1}{6}$&&\\&$\frac{1}{12}$&&$\frac{1}{12}$&\\$\frac{1}{6}$&&$\frac{1}{12}$&&
$\frac{1}{6}$\end{tabular}\\\\
$(N=3)$&&&&\begin{tabular}{ccccccc}&&&$\frac{1}{10}$&&&\\&&$\frac{1}{30}$&&$\frac{1}{30}$&&\\
&$\frac{1}{30}$&&$\frac{1}{60}$&&$\frac{1}{30}$&\\$\frac{1}{10}$&&$\frac{1}{30}$&&$\frac{1}{30}$&&$\frac{1}{10}$\end{tabular}
\end{tabular} 
&&&&
\begin{tabular}{lcccc}&&&&\\&&&&\\
$(N=4)$&&&&\begin{tabular}{ccccccccc}&&&&$\frac{1}{15}$&&&&\\&&&$\frac{1}{60}$&&$\frac{1}{60}$&&&\\&&$\frac{1}{90}$&&$\frac{1}{180}$&&$\frac{1}{90}$&&\\&$\frac{1}{60}$&&$\frac{1}{180}$&&$\frac{1}{180}$&&$\frac{1}{60}$&\\$\frac{1}{15}$&&$\frac{1}{60}$&&$\frac{1}{90}$&&$\frac{1}{60}$&&$\frac{1}{15}$\\\end{tabular}\\
&&&&\\
$(N=5)$&&&&\begin{tabular}{ccccccccccc}&&&&&$\frac{1}{21}$&&&&&\\&&&&$\frac{1}{105}$&&$\frac{1}{105}$&&&&\\&&&$\frac{1}{210}$&&$\frac{1}{420}$&&$\frac{1}{210}$&&&\\&&$\frac{1}{210}$&&$\frac{1}{630}$&&$\frac{1}{630}$&&$\frac{1}{210}$&&\\&$\frac{1}{105}$&&$\frac{1}{420}$&&$\frac{1}{630}$&&$\frac{1}{420}$&&$\frac{1}{105}$&\\$\frac{1}{21}$&&$\frac{1}{105}$&&$\frac{1}{210}$&&$\frac{1}{210}$&&$\frac{1}{105}$&&$\frac{1}{21}$\\
&&$\vdots$&&&$\vdots$&&&$\vdots$&&\end{tabular}
\end{tabular}
\end{tabular}
\end{center}  
The coefficients are given by 
\begin{equation}r^{(1)}_{N,n,m}=\frac{2}{(N+2)(N+1)\binom{N}{n\,,m}} \,.\label{tetraedro_Leibniz}\end{equation}
They satisfy the generalized Leibniz condition \eqref{generalized_Leibniz_rule}, and are defined in such a way that, when multiplying element by element by the Pascal pyramid, we get for the actual probabilities a uniform distribution:
\begin{equation}p^{(1)}_{N,n,m}\equiv\binom{N}{n\,,m}r^{(1)}_{N,n,m}=\frac{2}{(N+2)(N+1)}\end{equation}
with $\sum_{n+m=0}^Np^{(1)}_{N,n,m}=1$, which again leads as to a (bidimensional) $q-$Gaussian with $q \to -\infty$. 

As we did with the triangles, we can now get a family of scale-invariant pyramids out of the Leibniz-like pyramid. As the only layers of the pyramid with a central coefficient are those multiple of 3,  we shall descend 3 by 3 layers from the top of the Leibniz pyramid and divide the whole pyramid by the corresponding central element, that is, the element  $r^{(1)}_{3(\nu-1),\nu-1\nu-1}$ of the $3(\nu-1)-$th layer. This coefficient so turns to unity and we can start a new pyramid downwards from it, with coefficients given by  
\begin{equation}r^{(\nu)}_{N,n,m}=\frac{r^{(1)}_{N+3(\nu-1),n+\nu-1,m+\nu-1}}{{r^{(1)}_{3(\nu-1),\nu-1,\nu-1}}}.
\label{familia_piramides}\end{equation}
By construction, the family of pyramids \eqref{familia_piramides} satisfy the generalized Leibniz rule \eqref{generalized_Leibniz_rule} (see comment following Eq. \eqref{Leibniz_recursivo}), hence 
\begin{equation}r^{(\nu)}_{N,n,m}+r^{(\nu)}_{N,n+1,m-1}+r^{(\nu)}_{N,n,m-1}=r^{(\nu)}_{N-1,n,m-1}
\label{regla_tetraedro_nu}\end{equation}
for any positive integer $\nu$. Due to restriction \eqref{regla_tetraedro_nu} on the pyramid coefficients, the whole pyramid may be determined by only specifying the elements of one face.

Expressing the trinomial coefficient in \eqref{tetraedro_Leibniz} as a product of binomial coefficients and making use of the property of the Beta function expressed in \eqref{Leibniz_triangle}, the coefficients of the Leibniz-like pyramid can alternatively be expressed as
\begin{equation}r^{(\nu)}_{N,n,m}=\frac{2(N-n+1)}{N+2}B(N-n-m+1,m+1)B(N-n+1,n+1)\label{alternativa}\end{equation}
Introducing now \eqref{alternativa} in \eqref{familia_piramides}, after some algebra,  one finally gets
\begin{equation}r^{(\nu)}_{N,n,m}=\frac{B(N-n-m+\nu,n+m+2\nu)B(n+\nu,m+\nu)}{B(\nu,\nu)B(\nu,2\nu)};\quad\nu>0
\label{tetraedro_beta}\end{equation}
where now $\nu$ can take on any positive value.  

Multiplying the Pascal pyramid by the family of pyramids \eqref{tetraedro_beta}, one gets a family of probability distributions 
\begin{equation}p^{(\nu)}_{N,n,m}=\binom{N}{n,m}r^{(\nu)}_{N,n,m}\label{distribucion_generalizada}\end{equation}
with $\sum_{n+m=0}^Np^{(\nu)}_{N,n,m}=1$ for any $\nu$, associated to the new family of statistical variables 
\begin{equation}\vec X^{(\nu)}_{(2)}=\vec X^{(\nu)}_1+\vec X^{(\nu)}_2+\cdots+\vec X^{(\nu)}_N\label{vec_X_nu}\end{equation} 

\begin{figure}[h]

\begin{minipage}{.4\linewidth}
\centering\includegraphics[height=8cm,clip=]{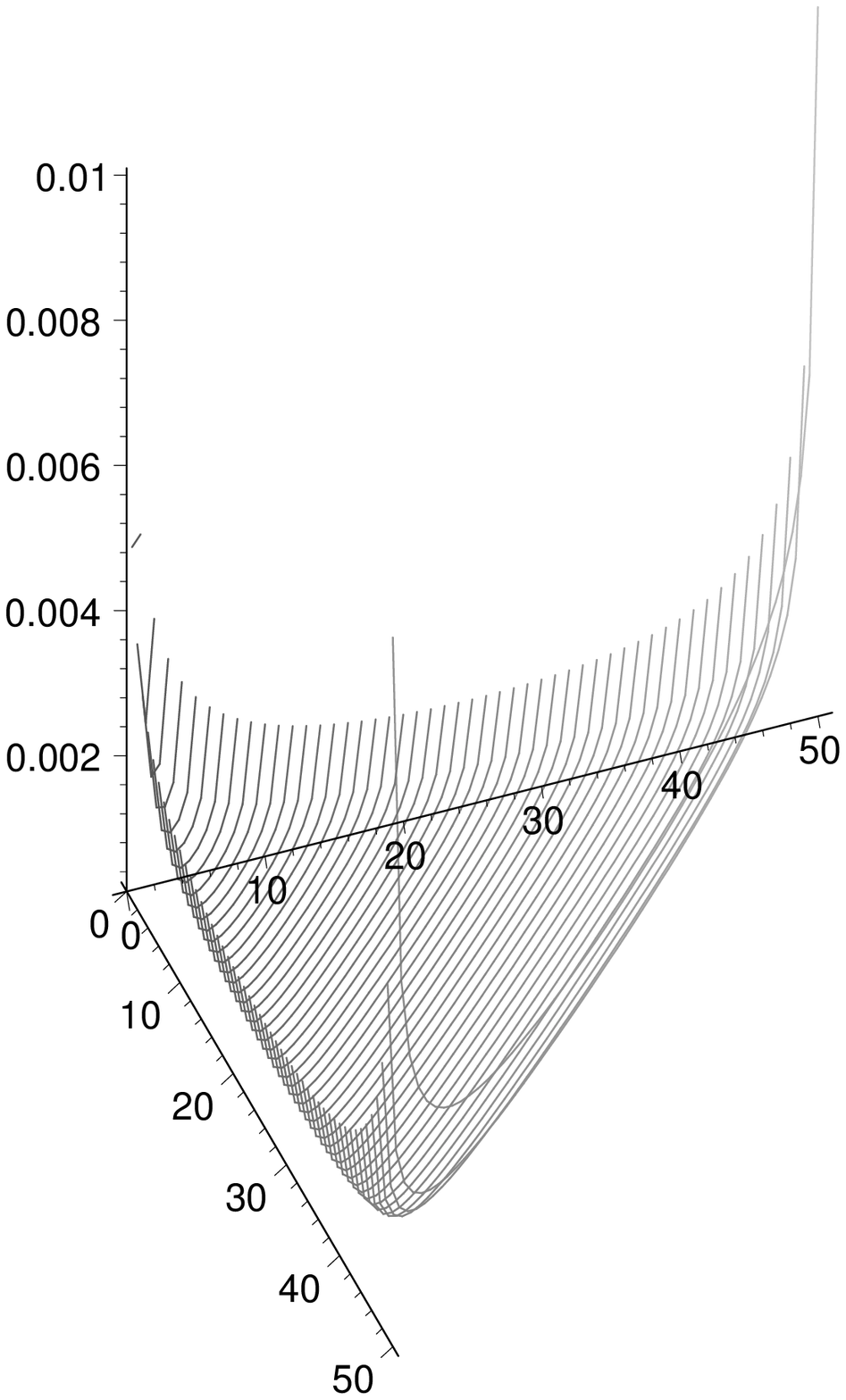}
\end{minipage}\hspace*{.5cm}
\begin{minipage}{.4\linewidth}
\centering\includegraphics[height=8cm,clip=]{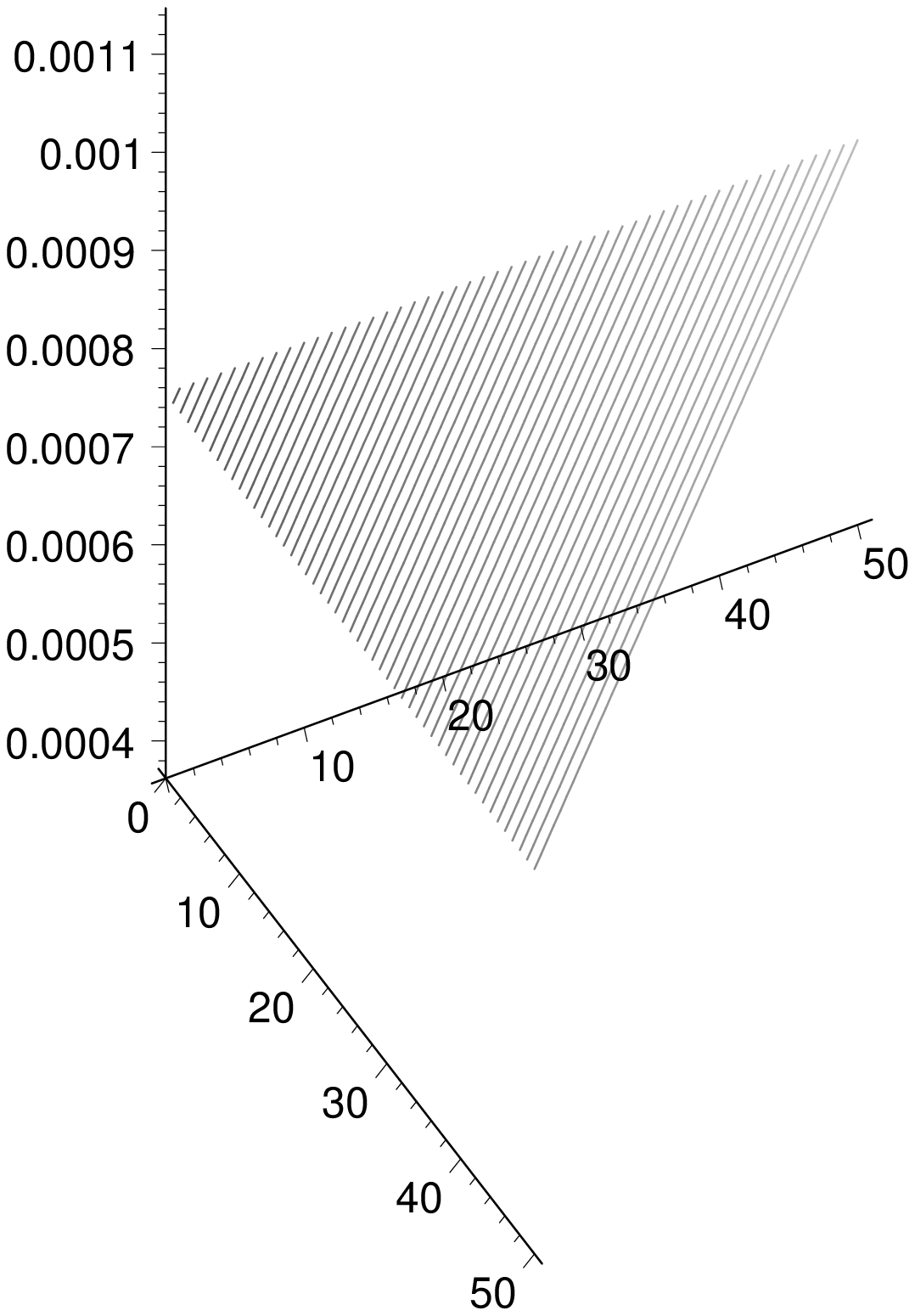}
\end{minipage}
\vspace*{3cm}
\begin{minipage}{.4\linewidth}
\centering\includegraphics[height=8cm,clip=]{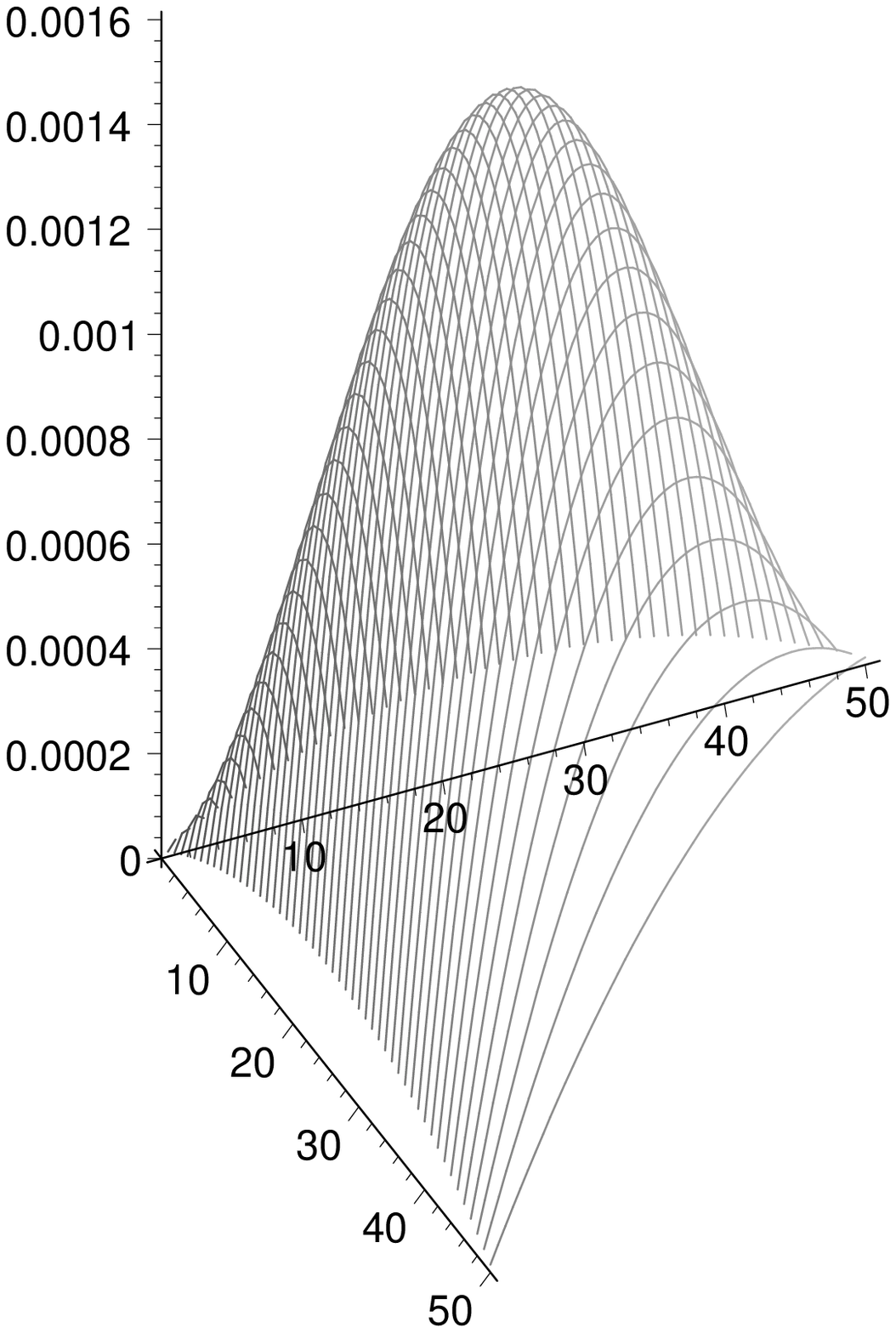}
\end{minipage}\hspace*{.5cm}
\begin{minipage}{.4\linewidth}
\centering\includegraphics[height=8cm,clip=]{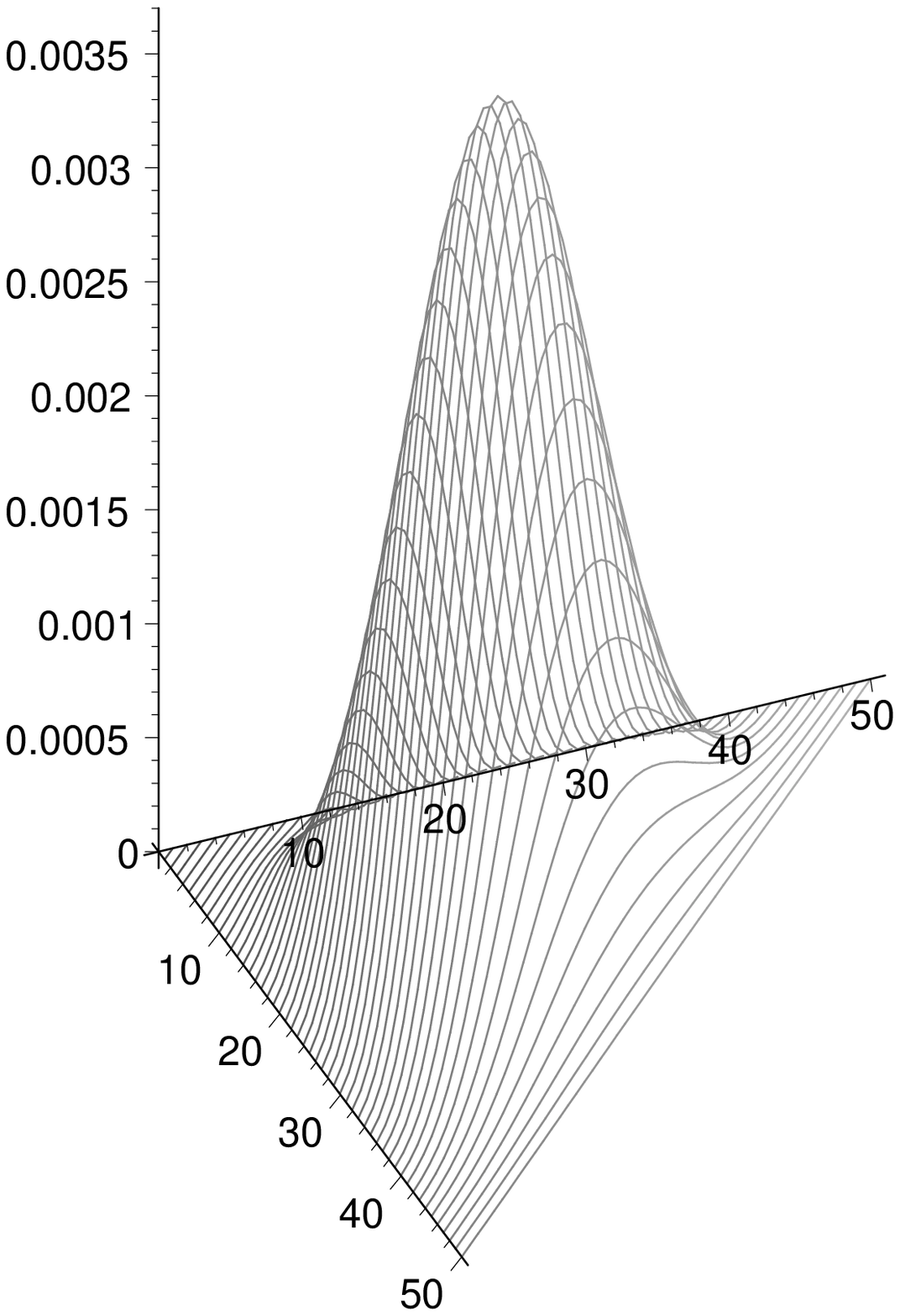}
\end{minipage}
\caption{Probability distributions \eqref{distribucion_generalizada} for $N=50$ and $\nu=\frac{1}{2}$ (top left), 1 (top right), 2 (bottom left) and 5 (bottom right).\label{probabilidades_2d}}
\end{figure}

Fig. \ref{probabilidades_2d} shows probability distributions \eqref{distribucion_generalizada} for $N=50$ and $\nu=\frac{1}{2}$, 1 (uniform distribution in the triangle $0\leqslant n+m\leqslant N$), 2 and 5. 

It may be shown that $\langle\vec X^{(\nu)}_{(2)}\rangle=\left(\frac{N}{3}, \frac{N}{3}\right)$ and the covariance matrix is 
\begin{equation}\Sigma^{(\nu)}_{(2)}=\frac{N(N+3\nu)}{9(1+3\nu)}\left(\begin{array}{cc}2&-1\\-1&2\end{array}\right)
\label{matriz_covarianzas}\end{equation}
Concerning the ternary variables $\vec X^{(\nu)}_i=(X^{(\nu)}_i,Y^{(\nu)}_i)$ in the sum \eqref{vec_X_nu} (which take the same values as the ternary independent variables $\vec X_i$ in \eqref{vec_X}), they are no longer independent of each other, their first and second order moments being given by
$\langle X^{(\nu)}_i\rangle=\langle Y^{(\nu)}_i\rangle=r^{(\nu)}_{1,0,0}=\frac{1}{3}\;\forall\nu$, $\sigma^2_{X^{(\nu)}_i}=\sigma^2_{Y^{(\nu)}_i}=\frac{2}{9}\;\forall\nu$,  $\sigma^2_{X^{(\nu)}_iX^{(\nu)}_j}=\sigma^2_{Y^{(\nu)}_iY^{(\nu)}_j}=r^{(\nu)}_{2,0,0}-\frac{1}{9}=\frac{2}{9(1+3\nu)}$ for $i\neq j$, $\sigma^2_{X_iY_i}=-\frac{1}{9}$ and $\sigma^2_{X^{(\nu)}_iY^{(\nu)}_j}=r^{(\nu)}_{2,1,0}-\frac{1}{9}=-\frac{1}{9(1+3\nu)}$ for $i\neq j$. The scale-invariant character of the correlations that we have introduce becomes, as before, apparent. 

\subsection{Boltzmann limit}
\label{Boltzmann_limit}
Taking limits in \eqref{tetraedro_beta} one gets $\lim_{\nu\to\infty}r^{(\nu)}_{N,n,m}=\frac{1}{3^N}\label{r_limite}$, so in the Boltzmann limit the trinomial distribution
\begin{equation}\lim_{\nu\to\infty}p^{(\nu)}_{N,n,m}=\binom{N}{n,m}\frac{1}{3^N}\label{p_limite}\end{equation}
with $p_1=p_2=1-p_1-p_2=\frac{1}{3}$ is obtained (as well as covariance matrix \eqref{matriz_covarianzas}, in the Boltzmann limit, coincides with covariance matrix \eqref{covarianzas} for the same values of $p$ and $q$), which is again consistent with the fact that $\lim_{\nu\to\infty}\sigma^2_{X^{(\nu)}_iX^{(\nu)}_j}=
\lim_{\nu\to\infty}\sigma^2_{Y^{(\nu)}_iY^{(\nu)}_j}=\lim_{\nu\to\infty}\sigma^2_{X^{(\nu)}_iY^{(\nu)}_j}=0$, so independent dices are recovered. Applying now the CLT, a Gaussian distribution appears again in the thermodynamic limit. 

We will devote Sec. \ref{joint-conditional-marginal} to the study of probability distribution \eqref{distribucion_generalizada}, in particular its thermodynamic limit, in order to check if it is a bidimensional $q-$Gaussian.  

\subsection{Entropy}

As it was shown in Ref. \cite{RodriguezSchwammleTsallis}, the family of distributions \eqref{probabilidad_nu} based on scale-invariant triangles \eqref{triangulos_nu}, are properly described by the Boltzmann-Gibbs entropy. It has been recently shown (see \cite{HanelThurner} and references therein) that this is so due to the fact the number of microstates increases exponentially with the system size ($\Omega(N)=2^N$). In turn, the nonadditive entropy $S_q$ with $q\neq 1$ \cite{Tsallis:88}  (the $q=1$ case corresponds to the Boltzmann entropy) 
% with $q=1-\frac{1}{b}$ (the $q=1$ case corresponds to the Boltzmann entropy) is
turns out to be appropriate (in the sense that the entropy is extensive) to describe systems for which a relevant fraction of the degrees of freedom vanishes in the thermodynamic limit \cite{HanelThurner}. Such is the case for the binary models studied in \cite{MarshFuentesMoyanoTsallis}, corresponding to triangles where the nonzero probabilities restrict, in the thermodynamic limit, to a strip of size $b$ on the triangle, being well described by the nonadditive entropy with $q=1-\frac{1}{b}$ (more precisely, for this value of $q$, $S_q$ is extensive).   

\begin{figure}\centering
\includegraphics[width=.8\linewidth,angle=-90]{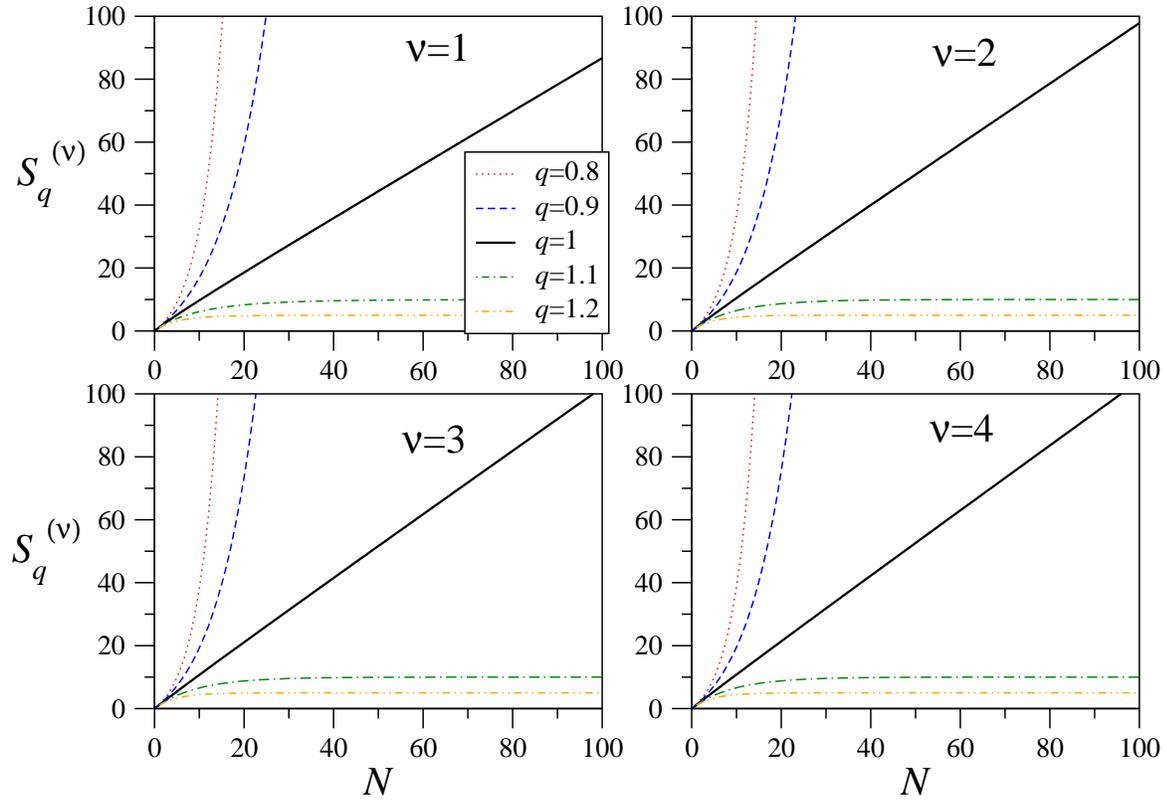}
\caption{(Color online) Tsallis entropy \eqref{entropia} for $q=$0.8, 0.9, 1, 1.1 and 1.2 of family \eqref{tetraedro_beta} for $\nu=1$, 2, 3 and 4. For all values of $\nu$ we have $q_\text{ent}=1$.\label{entropia_figura}}\end{figure}

For the ternary model \eqref{distribucion_generalizada} we also have exponential increase of the phase volume ($\Omega(N)=3^N$), so, once again, the Boltzmann entropy is expected to be extensive. The nonadditive entropy for the family of pyramids \eqref{tetraedro_beta} is given by
\begin{equation}S_q^{(\nu)}=\displaystyle\frac{1-\displaystyle\sum_{n+m=0}^N\binom{N}{n,m}\left(r^{(\nu)}_{N,n,m}\right)^q}{q-1}\label{entropia}\end{equation}
for $q\neq 1$ and $S_1^{(\nu)}=-\sum_{n+m=0}^N\binom{N}{n,m}r^{(\nu)}_{N,n,m}\ln\left(r^{(\nu)}_{N,n,m}\right)$ for $q=1$. Fig.~\ref{entropia_figura} shows   
$q-$entropy \eqref{entropia} as a function of $N$ for different values of $\nu$ and $q$. As predicted, independently of the value of $\nu$, the value of $q$ which makes the entropy $S_q^{(\nu)}$ extensive is $q_\text{ent}=1$.

\section{Joint, conditional and marginal distributions}
\label{joint-conditional-marginal}

In order to get a deeper insight into the two-dimensional probability distribution \eqref{distribucion_generalizada} we shall start by studying its associated one-dimensional marginal distributions since, as stated in Ref.~\cite{Vignat}, in case  these distributinos were $q-$Gaussians, the two-dimensional one would also be $q-$Gaussian, with  values of $q$ related by $\frac{2}{1-q_\text{1d}}=\frac{2}{1-q_\text{2d}}+1$, where $q_\text{1d}$ ($q_\text{2d}$) stands for the value of $q$ corresponding to the one- (two-)dimensional $q-$Gaussian. 

\begin{figure}
\centering\includegraphics[width=\linewidth,clip=]{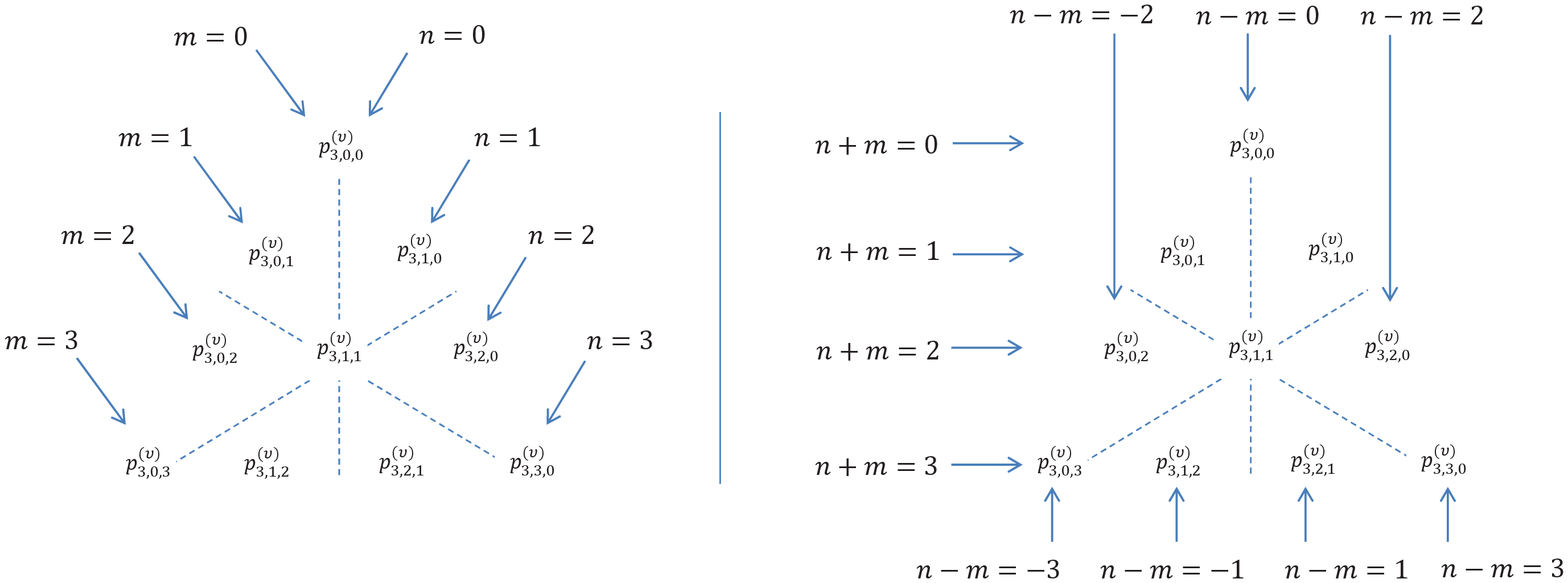}\caption{$N=3$ layer of pyramid  \eqref{distribucion_generalizada}. The three heights of the triangle are shown. As the elements of the triangle are symmetric with respect to any of them, there are only three different elements in this layer. Left: ascending (descending) diagonals of the layer correspond to constant values of $n$ ($m$). Right: rows (columns) of the layer correspond to constant values of $n+m$ ($n-m$). \label{figura_marginales}}\end{figure}

\begin{figure}
\centering\includegraphics[height=\linewidth,angle=-90,clip=]{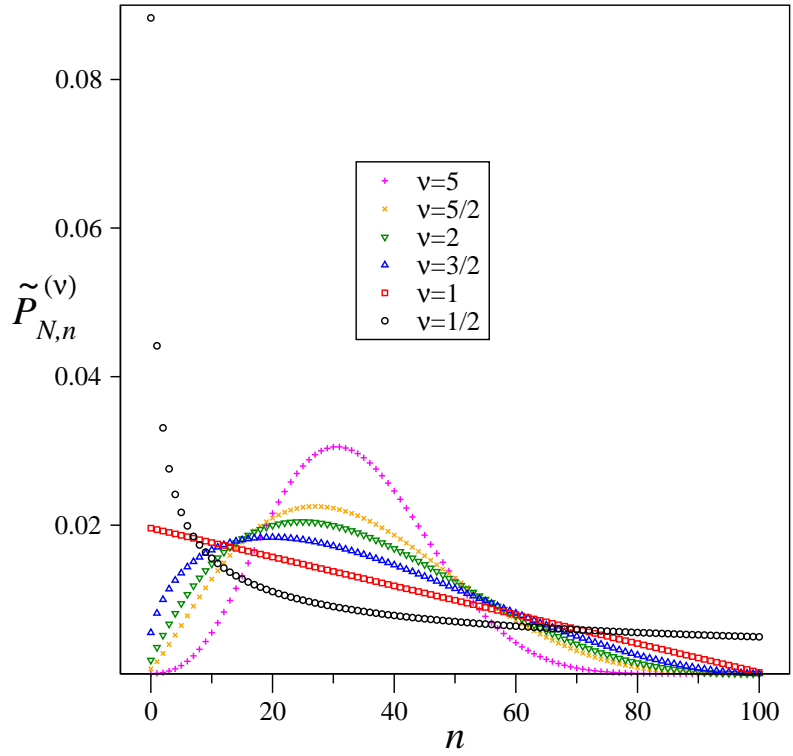}
\caption{(Color online) Marginal probability distributions $\tilde p^{(\nu)}_{N,n}$ for $N=100$ and $\nu=\frac{1}{2}$, 1, $\frac{3}{2}$, 2, $\frac{5}{2}$ and 5.\label{marginales}}
\end{figure}

By construction, probability distribution \eqref{distribucion_generalizada} inherits the highly symmetrical character of the trinomial coefficients (any layer of the Pascal pyramid is a triangle symmetric with respect to any of its heights) and so $p^{(\nu)}_{N,n,m}=p^{(\nu)}_{N,m,n}=p^{(\nu)}_{N,N-n-m,m}=p^{(\nu)}_{N,n,N-n-m}=
p^{(\nu)}_{N,m,N-n-m}=p^{(\nu)}_{N,n,N-n-m}$ (the same property holds for coefficients \eqref{tetraedro_beta}). Thus the marginal distributions $\tilde p^{(\nu)}_{N,n}\equiv\sum_{m=0}^{N-n}p^{(\nu)}_{N,n,m}$, $\tilde p^{(\nu)}_{N,m}\equiv\sum_{n=0}^{N-m}p^{(\nu)}_{N,n,m}$ and $\tilde p^{(\nu)}_{N,l}\equiv\sum_{n+m=l}p^{(\nu)}_{N,n,m}$, associated to variables $n$, $m$ and $l=n+m\in\{0,1,\dots,N\}$  (which, as shown in Fig.~\ref{figura_marginales},  correspond respectively to adding up probabilities along the ascending diagonals, descending diagonals or rows  of the layers of the pyramid) are identical. Fig.~\ref{marginales} shows probability distributions $\tilde p^{(\nu)}_{N,n}$ for $N=100$ and $\nu=1/2$, 1, $\frac{3}{2}$, 2, $\frac{5}{2}$ and 5. The first apparent property of these distributions is its {\em non}symmetric character, their first moments being given by $\langle X\rangle=\frac{N}{3}$ and, as easily deduced from \eqref{matriz_covarianzas}, $\sigma^2_X=\frac{2N(N+3\nu)}{9(1+3\nu)}$. This asymmetry vanishes when $\nu\to\infty$ (as stated in Sec. \ref{Boltzmann_limit}, in the Boltzmann limit, bidimensional distribution \eqref{distribucion_generalizada} tends to a symmetric trinomial distribution, thus having symmetric binomial marginal distributions) but remains when increasing $N$ (it may be shown that the skewness coefficient assymptotically approaches a finite value in the thermodynamic limit).  

A new family of nonsymmetric triangles can be defined from the marginal distributions $\tilde p^{(\nu)}_{N,n}$ in the form 
\begin{equation}\tilde r_{N,n}^{(\nu)}=\frac{\tilde p_{N,n}^{(\nu)}}{\binom{N}{n}}\label{triangulos_no_simetricos}\end{equation}
The $\nu=1$ instance of triangles \eqref{triangulos_no_simetricos} is
    
\begin{center}\begin{tabular}{ccccccccccc}&&&&&$\tilde r^{(1)}_{N,n}$&&&&&\\&&&&&&&&&&\\&&&&&$1$&&&&&\\&&&&$\frac{2}{3}$&&$\frac{1}{3}$&&&&\\&&&$\frac{1}{2}$&&$\frac{1}{3}$&&$\frac{1}{6}$&&&\\&&$\frac{2}{5}$&&$\frac{3}{10}$&&$\frac{1}{5}$&&$\frac{1}{10}$&&\\&$\frac{1}{3}$&&$\frac{4}{15}$&&$\frac{1}{5}$&&$\frac{2}{15}$&&$\frac{1}{15}$&\\$\frac{2}{7}$&&$\frac{5}{21}$&&$\frac{4}{21}$&&$\frac{1}{7}$&&$\frac{2}{21}$&&$\frac{1}{21}$\\&&$\vdots$&&&&&&$\vdots$&&\end{tabular}\end{center}

Though asymmetric, triangles \eqref{triangulos_no_simetricos} fulfill scale-invariance condition \eqref{Leibniz_rule_nu}. In fact, they can alternatively be obtained applying rule \eqref{Leibniz_rule_nu} to the edge $r^{(\nu)}_{N,0,0}=\tilde r^{(\nu)}_{N,N}$ of the corresponding pyramid. 

In addition, for integer values of $\nu$, triangles \eqref{triangulos_no_simetricos} may be obtained as nonsymmetric, properly rescaled subtriangles of the family \eqref{triangulos_nu} in the form 
\begin{equation}\tilde r^{(\nu)}_{N,n}=\frac{r^{(\nu)}_{N+\nu,n}}{r^{(\nu)}_{\nu,0}}\label{triangulos_asimetricos_recursivo}\end{equation}
From \eqref{triangulos_asimetricos_recursivo}, the marginal probabilities $\tilde p_{N,n}$ can be related to the one-dimensional family \eqref{probabilidad_nu} in the form 
\begin{equation}\tilde p^{(\nu)}_{N,n}=\frac{1}{r^{(\nu)}_{\nu,0}}\frac{(N-n+1)\cdots(N-n+\nu)}{(N+1)\cdots(N+\nu)}p^{(\nu)}_{N+\nu,n},\label{probabilidades_marginales}\end{equation}
which, for $\nu=1$, reduces to the straight line 
\begin{equation}\tilde p^{(1)}_{N,n}=\frac{2}{(N+1)(N+2)}(N-n+1)\label{recta}\end{equation}
shown in Fig.~\ref{marginales}.

\begin{figure}
\centering\includegraphics[height=\linewidth,angle=-90,clip=]{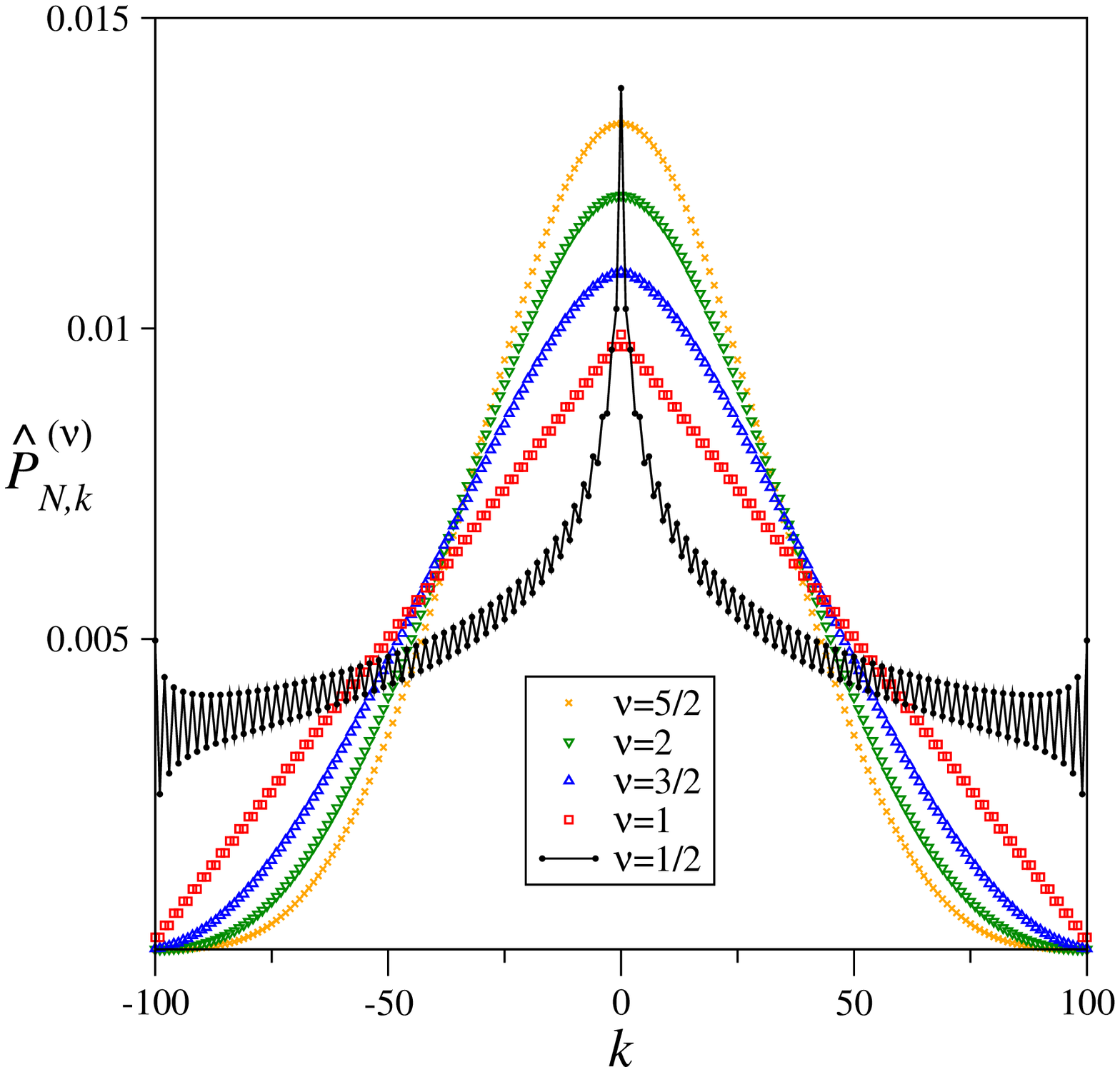}
\caption{(Color online) Marginal probability distributions $\hat p^{(\nu)}_{N,k}$ for $N=100$ and $\nu=\frac{1}{2}$, 1, $\frac{3}{2}$, 2 and $\frac{5}{2}$.   \label{marginales_n-m}}
\end{figure}

The nonsymmetric character of marginal distributions $\tilde p^{(\nu)}_{N,n}$ makes them not good candiates to yield $q-$Gaussians in the thermodynamic limit. Nevertheless, there is another yet unexplored direction for the marginal distributions. We shall now add up coefficients in the vertical direction of each layer, which corresponds to calculate the marginal distribution $\hat p_{N,k}^{(\nu)}\equiv\sum_{n-m=k}p^{(\nu)}_{N,n,m}$ associated to the variable $k=n-m\in\{-N,\dots,0,\dots,N\}$ (see Fig.~\ref{figura_marginales}). Fig.~\ref{marginales_n-m} shows marginal distributions $\hat p_{N,k}^{(\nu)}$ for $N=100$ and $\nu=\frac{1}{2}$, 1, $\frac{3}{2}$, 2 and $\frac{5}{2}$. Contrary to the former case, these new marginal distributions are symmetric, due to the aforementioned fact that the layers of the pyramid are symmetric triangles with respect to their heights. From these distributions and dividing by the appropiate binomial coefficient, we can now define a new family of triangles in the form 
\begin{equation}\hat r_{2N,N+k}^{(\nu)}=\frac{\hat p_{N,k}^{(\nu)}}{\binom{2N}{N+k}},\quad k=-N,\dots,0,\dots,N\label{triangulos_simetricos}\end{equation}
with only even labelled rows. The $\nu=2$ instance of triangles \eqref{triangulos_simetricos} is
\begin{center}
\begin{tabular}{ccccccccccccc}&&&&$\hat r^{(2)}_{2N,N+k}$&&&\\&\\&&&&$1$&&&\\&\\&&&$\frac{1}{3}$&$\frac{1}{6}$&$\frac{1}{3}$&&\\&\\&&$\frac{1}{7}$&$\frac{1}{21}$&$\frac{1}{18}$&$\frac{1}{21}$&$\frac{1}{7}$&\\&\\&$\frac{1}{14}$&$\frac{1}{56}$&$\frac{1}{70}$&$\frac{3}{280}$&$\frac{1}{70}$&$\frac{1}{56}$&$\frac{1}{14}$\\&\\$\frac{5}{126}$&$\frac{1}{126}$&$\frac{17}{3528}$&$\frac{5}{1764}$&$\frac{13}{4410}$&$\frac{5}{1764}$&$\frac{17}{3528}$&$\frac{1}{126}$&$\frac{5}{126}$\\&$\vdots$&&&&&&$\vdots$&&&&
\end{tabular}
\end{center}

\begin{figure}\centering
\includegraphics[height=\linewidth,angle=-90,clip=]{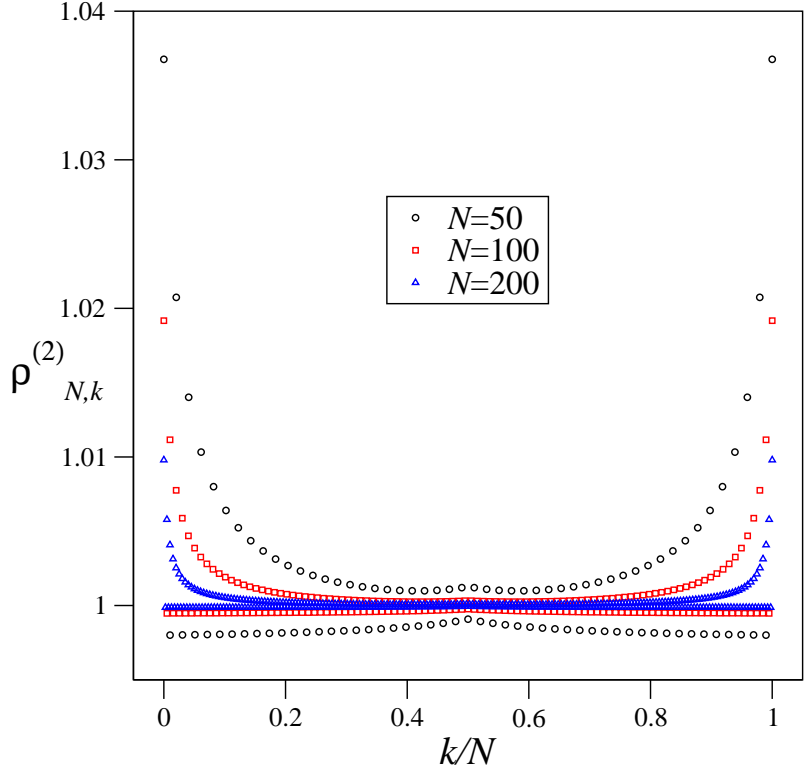}
\caption{(Color online) Quotient \eqref{cociente} versus $k/N$ for $\nu=2$ and $N=50$, $100$ and $200$. When increasing $N$, $\rho^{(2)}_{N,k}$ approaches 1.\label{cociente_figura}}
\end{figure}

Though symmetric triangles \eqref{triangulos_simetricos} share their sides with one of the sides of triangles \eqref{triangulos_no_simetricos} and the edge of pyramids \eqref{tetraedro_beta}, i.e., $\hat r^{(\nu)}_{2N,0}=\tilde r^{(\nu)}_{N,N}=r^{(\nu)}_{N,0,0}$, it should be stressed that they are {\em not} scale invariant. To properly define scale invariance in triangles \eqref{triangulos_simetricos}, Leibniz rule \eqref{Leibniz_rule} can recursively be applied to connect coefficients separated two rows, thus obtaining $\hat r^{(\nu)}_{2N,k}+2\hat r^{(\nu)}_{2N,k+1}+\hat r^{(\nu)}_{2N,k+2}=\hat r^{(\nu)}_{2N-2,k}$.  This relation is {\em not} fulfilled by  triangles \eqref{triangulos_simetricos}. Nevertheless, it is asymptotically fulfilled in the thermodynamic limit. In effect, we shall define the quotient
\begin{equation}\rho^{(\nu)}_{N,k}=\frac{\hat r^{(\nu)}_{2N-2,k}}{\hat r^{(\nu)}_{2N,k}+2\hat r^{(\nu)}_{2N,k+1}+\hat r^{(\nu)}_{2N,k+2}}\label{cociente}\end{equation}
and show that $\lim_{N\to\infty}\rho^{(\nu)}_{N,n}=1$. Fig.~\ref{cociente_figura} shows quotient \eqref{cociente} versus $k/N$ for $\nu=2$ and $N=50$, $100$ and $200$. An oscillating trend around the value 1 is observed, the quotient being closer to 1 for the central values of $k$. Increasing $N$ makes $\rho^{(2)}_{N,k}$ closer to 1. We then verify that triangles \eqref{triangulos_simetricos} are {\em asymptotically scale invariant.} 

There is, however, an alternative way of obtaining the same above distributions $\hat p_{N,k}^{(\nu)}$. Let us think of our ternary variables as being scalars ---instead of vectors, as imposed in \eqref{vec_X_nu}--- having values $-1$, 0 and 1. We shall denote them by $\xi_i^{(\nu)}$, $i=1,\dots,N$, and consider the sum of them $\xi^{(\nu)}=\sum_{i=1}^N\xi_i^{(\nu)}$, which is a scalar variable having values in $\{-N,\dots,0,\dots,N\}$. If we associate the probabilities of pyramid \eqref{distribucion_generalizada} to the probability of having $n$ variables with value 1 and $m$ variables with value $-1$, the probability distribution of variable $\xi^{(\nu)}$ exactly coincides with the marginal distribution $\hat p_{N,k}^{(\nu)}$ of distribution \eqref{distribucion_generalizada}. Thus, in our model the dimension associated to the random variables plays a secondary role, the specific form of the correlations among them being of major importance. Some numerical results seem to indicate that marginal distributions $\hat p_{N,k}^{(\nu)}$ do not yield $q-$Gaussians in the thermodynamic limit either. Nevertheless, the difficulty to reach high values of $N$ (due to the rapidly increasing value of the trinomial coefficients) doesn't allow us to establish a definite conclusion.

Let us turn now to the study of the conditional distributions of probability distribution \eqref{distribucion_generalizada}, which are defined in the way
\begin{equation}p^{(\nu)}_{N,n|m}=\frac{p^{(\nu)}_{N,n,m}}{\tilde p^{(\nu)}_{N,m}};\quad n=0,\dots,N-m\label{condicionales}\end{equation}
and an equivalent expression for $p^{(\nu)}_{N,m|n}$, which, for symmetry considerations, is identical. We show in the Appendix that 
\begin{equation}p^{(\nu)}_{N,n|m}=p^{(\nu)}_{N-m,n}\label{condicionales_relacion}\end{equation}
i.e., the conditional distributions of the family of distributions \eqref{distribucion_generalizada} coincide with the family \eqref{probabilidad_nu}, already shown in Fig. \ref{probabilidades_1d}, for an appropiate system size. 

Thus, bidimensional distribution \eqref{distribucion_generalizada} contains sections which, as stated in Sec. \ref{Boltzmann_1d}, are $q-$Gaussians in the thermodynamic limit, which is a necessary, but by no means sufficient condition for probability distributions \eqref{distribucion_generalizada} to be two-dimensional $q-$Gaussians.  

To further explore the nature of distribution \eqref{distribucion_generalizada} we will do the following. Let us first typify variable $\vec X^{(\nu)}_{(2)}$ by making the linear change 
\begin{equation}\left(\begin{array}{c}n\\m\end{array}\right)\to\left(\begin{array}{c}u\\v\end{array}\right)=A\left(\begin{array}{c}n-N/3\\m-N/3\end{array}\right)\label{cambio_2d}\end{equation}
with jacobian $(\det A)^{-1}$, where matrix $A$ is defined in such a way that $(\Sigma^{(\nu)}_{(2)})^{-1}=A^TA$, with $\Sigma^{(\nu)}_{(2)}$ given in \eqref{matriz_covarianzas}, thus 
\begin{equation}A=\sqrt{\frac{3(1+3\nu)}{2N(N+3\nu)}}\left(\begin{array}{cc}\sqrt{3}&\sqrt{3}\\-1&1\end{array}\right)\label{linear_change}\end{equation}
Vector random variable $\vec U=(u,v)^T$ defined in \eqref{cambio_2d} is centered, i.e., $E[\vec U]=A(E[\vec X^{(\nu)}_{(2)}]-\vec \mu)=\vec 0$, where $\vec\mu\equiv\left(\frac{N}{3},\frac{N}{3}\right)^T$, and has identity covariance matrix, since $E[\vec U\vec U^T]=E[A(\vec X^{(\nu)}_{(2)}-\vec \mu)(\vec X^{(\nu)}_{(2)}-\vec\mu)^TA^T]=AE[(\vec X^{(\nu)}_{(2)}-\vec \mu)(\vec X^{(\nu)}_{(2)}-\vec\mu)^T]A^T=A\Sigma^{(\nu)}_{(2)} A^T=I$.
In addition, change \eqref{cambio_2d} has the extra effect of converting triangle $0\leqslant n+m\leqslant N$ of vertices $(0,0)$, $(N,0)$ and $(0,N)$, into a centered in the origin and equilateral triangle of vertices $\alpha(-1,0)$, $\alpha(1/2,-\sqrt{3}/2)$ and $\alpha(1/2,\sqrt{3}/2)$, with $\alpha=\sqrt{\frac{2(1+3\nu)}{N(N+3\nu)}}$, more appropiate for our purpose.

\begin{figure}
\begin{minipage}{.3\linewidth}
\centering\includegraphics[width=4.5cm]{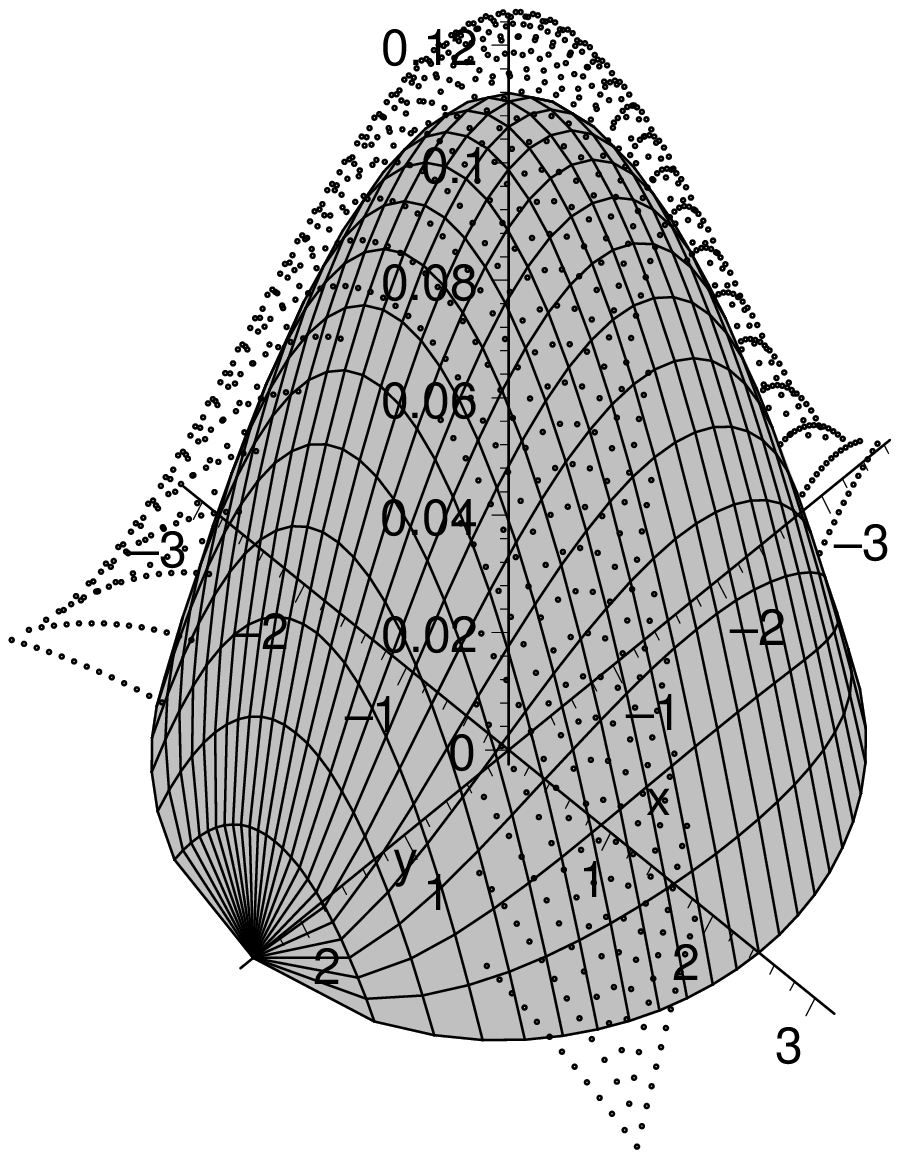}\end{minipage}
\hspace*{.25cm}
\begin{minipage}{.3\linewidth}
\centering\includegraphics[width=4.5cm]{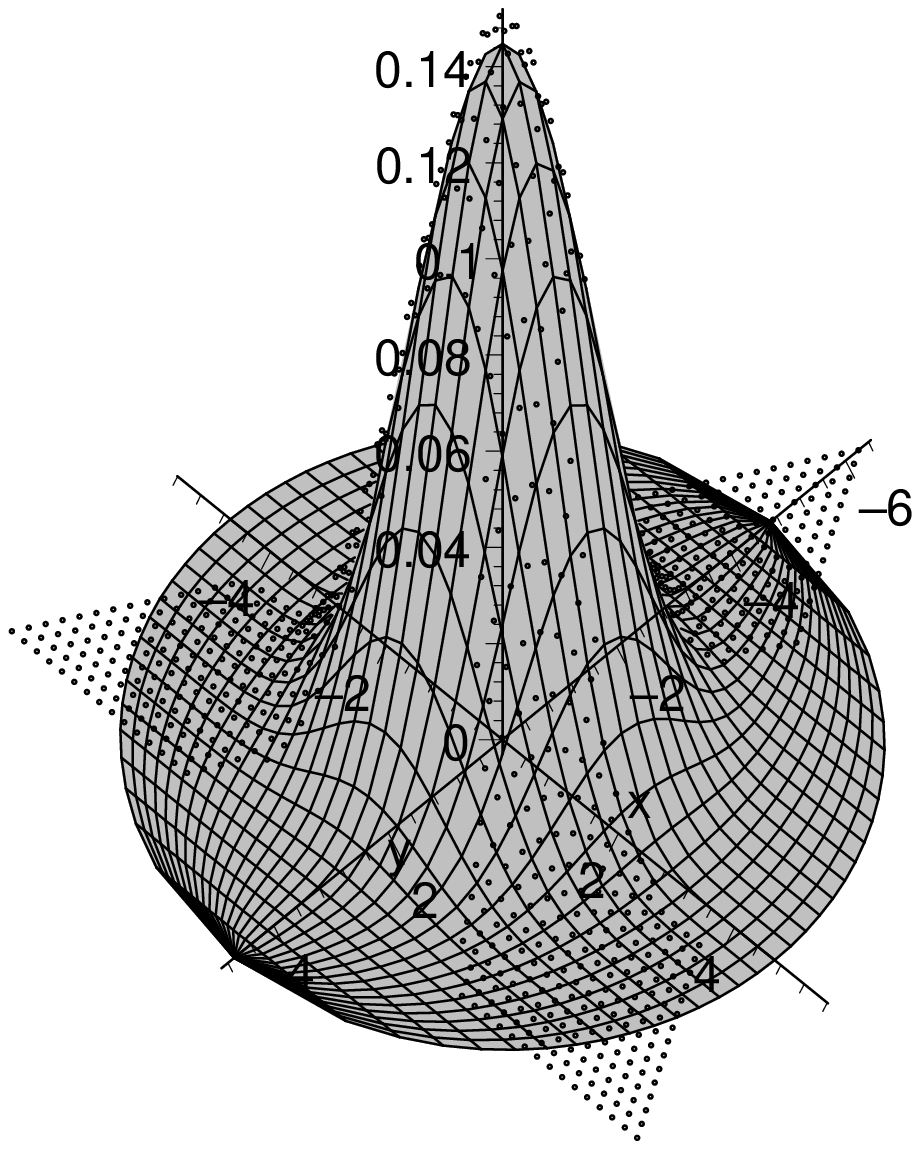}\end{minipage}
\hspace*{.25cm}
\begin{minipage}{.3\linewidth}
\centering\includegraphics[width=4.5cm]{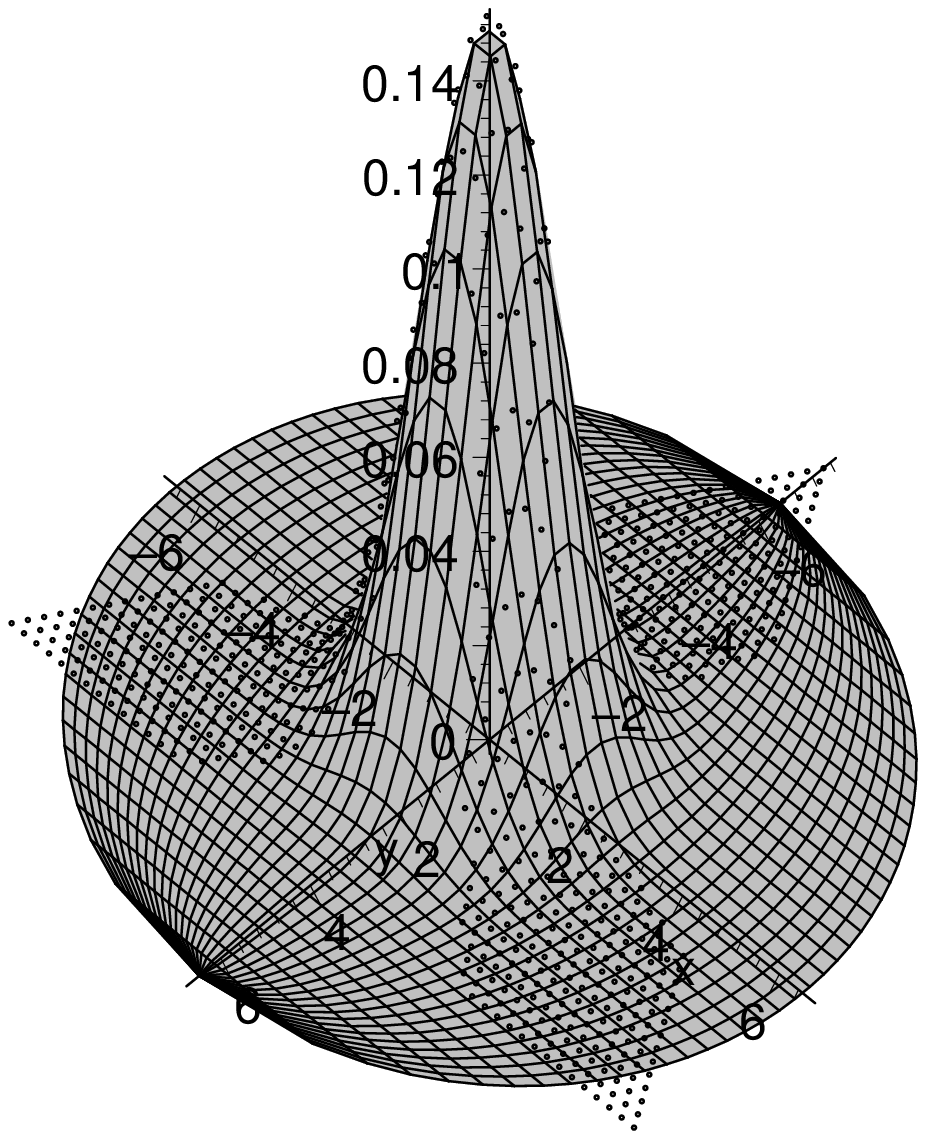}
\end{minipage}\caption{$(\det A)^{-1}p^{(\nu)}_{N,n,m}$ versus $\vec U$ (dots) compared with the corresponding bidimensional $q_\nu$-Gaussian (solid surface) for $N=50$ and  $\nu=2$, $q_\nu=0$ (left), $\nu=10$, $q_\nu=\frac{8}{9}$ (center) and  $\nu=20$, $q_\nu=\frac{18}{19}$ (right). \label{comparacion_figura}}\end{figure}

In the hypothesis that distribution \eqref{distribucion_generalizada} yields a bidimensional $q_\nu-$Gaussian with $q_\nu=\frac{\nu-2}{\nu-1}$ in the thermodynamic limit, we shall compare probability distribution of variable $\vec U$ with a typified bidimensional $q_\nu-$Gaussian \eqref{q-gaussiana_d} with $\beta=\frac{1}{6-4q_\nu}$ and $\Sigma=I$ (thus having identity covariance matrix as ensured by Eq.~\eqref{covarianzas_q-gaussiana_d}). 
Fig.~\ref{comparacion_figura} plots $(\det A)^{-1}p^{(\nu)}_{N,n,m}$ versus $\vec U$ (dots) compared with the corresponding bidimensional $q_\nu$-Gaussian (solid surface), for $N=50$ and  $\nu=2$, $q_\nu=0$ (left), $\nu=10$, $q_\nu=\frac{8}{9}$ (center) and  $\nu=20$, $q_\nu=\frac{18}{19}$ (right). For $\nu=2$ the mismatch is evident, the results not improving when increasing $N$, neither by changing the value of $q$. Nevertheless, when increasing $\nu$ the fitting improves, which is due to the fact that, as stated in Sec. \ref{Boltzmann_limit}, family \eqref{distribucion_generalizada} approaches a trinomial distribution in the Boltzmann limit, which in turn approaches a Gaussian distribution in the thermodynamic limit, while the corresponding $q_\nu-$Gaussian also approaches a Gaussian since $\lim_{\nu\to\infty}q_\nu=1$.  

Thus, bidimensional $q-$Gaussians with generic values of $q$ are elusive in the present model. A possible reason for it is the inadequacy of linear change \eqref{cambio_2d}. Instead, a nonlinear (and highly not trivial) change is needed, which transforms the triangle $0\leqslant m+n\leqslant N$ into a circle (either for all values of $N$, or at least for increasing $N$). Research along this line is in progress.     

\section{Generalization of the model to arbitrary dimension}
\label{dimension_arbitraria}
We may now throw a four-sided dice (i.e., a tetrahedric dice), with associated probabilities $p_1$, $p_2$, $p_3$ (with $p_1+p_2+p_3<1$), and $1-p_1-p_2-p_3$ for the different sides. We need now a three components variable 
\begin{equation}\vec X_{(3)}=\vec X_1+\vec X_2+\cdots+\vec X_N\end{equation} 
with $\vec X_{(3)}=(X,Y,Z)$, defined as a sum of $N$ quaternary variables $\vec X_i=(X_i,Y_i,Z_i)$, for $i=1,\dots,N$, taking the four possible values $(1,0,0)$, $(0,1,0)$, $(0,0,1)$ and $(0,0,0)$ associated to the different sides, so that $X$, $Y$ and $Z$ count the number of appearences of three of the sides out of $N$ throwings. Variable $\vec X_{(3)}$ follows the tetranomial distribution
\begin{equation}P(X=n,Y=m, Z=l)=\binom{N}{n,m,l}p_1^np_2^mp_3^l(1-p_1-p_2-p_3)^{N-n-m-l}\label{tetranomial}\end{equation}
where $0\leqslant n+m+l\leqslant N$ and the tetranomial coefficients,  $\binom{N}{n,m,l}=\binom{N}{n}\binom{N-n}{m}\binom{N-n-m}{l}$, stand for the number of ways in which the same result $(X,Y,Z)=(n,m,l)$ can be obtained after $N$ throwings. We have now $\Omega(N)=4^N$ different events, though only $\frac{(N+1)(N+2)(N+3)}{6}$ different probability values, namely $r_{N,n,m,l}\equiv p_1^np_2^mp_3^l(1-p_1-p_2-p_3)^{N-n-m-l}$. In order to display them (what we will not do) we would need to use a $\frac{(N+1)(N+2)(N+3)}{6}$ elements pyramid for each value of $N$ (in the same way as we used a $\frac{(N+1)(N+2)}{2}$ elements triangle for each value of $N$ for the trinomial distribution and a $N+1$ elements row for each value of $N$ for the binomial distribution). Putting all the pyramids together makes a {\em hyperpyramid} (in the same way as we got a pyramid made of triangles for the trinomial distribution and a triangle made of rows for the binomial distribution). In order to get the actual probabilities \eqref{tetranomial}, this hyperpyramid should be multiplied by what we may call {\em Pascal hyperpyramid} 

%\begin{scriptsize}
\begin{tabular}{ccccccccccccc}$(N=0)$&&&$(N=1)$&&&$(N=2)$&&&$(N=3)$&&$(N=4)$&$\cdots$
\\&&&&&&&&&&&&\\
\begin{tabular}{c}1\end{tabular}&&&
\begin{tabular}{ccc}&1&\\\\&1&\\1&&1\end{tabular}&&&
\begin{tabular}{ccccc}&&1&&\\\\&&2&&\\&2&&2&\\\\&&1&&\\&2&&2&\\1&&2&&1\end{tabular}&&&
\begin{tabular}{ccccccc}&&&1&&&\\\\&&&3&&&\\&&3&&3&&\\\\&&&3&&&\\&&6&&6&&\\&3&&6&&3&\\\\&&&1&&&\\&&3&&3&&\\&3&&6&&3&\\1&&3&&3&&1\end{tabular}&&
\begin{tabular}{ccccccccc}&&&&1&&&&\\\\&&&&4&&&&\\&&&4&&4&&&\\\\&&&&6&&&&\\&&&12&&12&&&\\&&6&&12&&6&&\\\\&&&&4&&&&\\&&&12&&12&&&\\&&12&&24&&12&&\\&4&&12&&12&&4&\\\\&&&&1&&&&\\&&&4&&4&&&\\&&6&&12&&6&&\\&4&&12&&12&&4&\\1&&4&&6&&4&&1\\\end{tabular}\end{tabular}
%\end{scriptsize}

\noindent which displays the tetranomial coefficients. Now, layers from top to bottom within $N-$th pyramid, correspond to constant values of $n+m+l$ from 0 to $N$. In addition, rows from bottom to top within each layer correspond to constant values of $l$ from 0 to $l+m+n$, while indexes $m$ and $n$ behave in the same way as in the Leibniz pyramid.  

We may define now the {\em Leibniz hyperpyramid} as 
\begin{equation}r^{(1)}_{N,n,m,l}=\frac{6}{(N+1)(N+2)(N+3)\binom{N}{n,l,m}}
\label{hiperpiramide_Leibniz}\end{equation}
whose coefficients may be displayed as 

\begin{scriptsize}
\begin{flushleft}
\begin{tabular}{cccccc}
$(N=0)$&$(N=1)$&$(N=2)$&$(N=3)$&$(N=4)$&$\cdots$\\
&\hspace*{-.2cm}&&&\\
\begin{tabular}{c}1\end{tabular}&\begin{tabular}{ccc}&$\frac{1}{4}$&\\\\&$\frac{1}{4}$&\\$\frac{1}{4}$&&$\frac{1}{4}$
\end{tabular}&
\begin{tabular}{ccccc}&&$\frac{1}{10}$&&\\\\
&&$\frac{1}{20}$&&\\&$\frac{1}{20}$&&$\frac{1}{20}$&\\\\
&&$\frac{1}{10}$&&\\&$\frac{1}{20}$&&$\frac{1}{20}$&\\$\frac{1}{10}$&&$\frac{1}{20}$&&$\frac{1}{10}$\end{tabular}&
\begin{tabular}{ccccccc}&&&$\frac{1}{20}$&&&\\\\
&&&$\frac{1}{60}$&&&\\&&$\frac{1}{60}$&&$\frac{1}{60}$&&\\\\
&&&$\frac{1}{60}$&&&\\&&$\frac{1}{120}$&&$\frac{1}{120}$&&\\&$\frac{1}{60}$&&$\frac{1}{120}$&&$\frac{1}{60}$&\\\\
&&&$\frac{1}{20}$&&&\\&&$\frac{1}{60}$&&$\frac{1}{60}$&&\\&$\frac{1}{60}$&&$\frac{1}{120}$&&$\frac{1}{60}$&\\$\frac{1}{20}$&&$\frac{1}{60}$&&$\frac{1}{60}$&&$\frac{1}{20}$\end{tabular}&
\begin{tabular}{ccccccccc}&&&&$\frac{1}{35}$&&&&\\\\&&&&$\frac{1}{140}$&&&&\\&&&$\frac{1}{140}$&&$\frac{1}{140}$&&&
\\\\&&&&$\frac{1}{210}$&&&&\\&&&$\frac{1}{420}$&&$\frac{1}{420}$&&&\\&&$\frac{1}{210}$&&$\frac{1}{420}$&&$\frac{1}{210}$&&\\\\
&&&&$\frac{1}{140}$&&&&\\&&&$\frac{1}{420}$&&$\frac{1}{420}$&&&\\&&$\frac{1}{420}$&&$\frac{1}{840}$&&$\frac{1}{420}$&&
\\&$\frac{1}{140}$&&$\frac{1}{420}$&&$\frac{1}{420}$&&$\frac{1}{140}$&\\\\
&&&&$\frac{1}{35}$&&&&\\&&&$\frac{1}{140}$&&$\frac{1}{140}$&&&\\&&$\frac{1}{210}$&&$\frac{1}{420}$&&$\frac{1}{210}$&&\\&$\frac{1}{140}$&&$\frac{1}{420}$&&$\frac{1}{420}$&&$\frac{1}{140}$&\\$\frac{1}{35}$&&$\frac{1}{140}$&&$\frac{1}{210}$&&$\frac{1}{140}$&&$\frac{1}{35}$\\\end{tabular}\end{tabular}
\end{flushleft}
\end{scriptsize}

So defined, when multiplying Pascal hyperpyramid by Leibniz hyperpyramid one obtains the uniform tridimensional distribution given by

\begin{equation}p^{(1)}_{N,n,m,l}=\frac{6}{(N+1)(N+2)(N+3)}\label{hiperpiramide_Leibniz_probabilidad}\end{equation}

In adittion, a family of hyperpyramids may be extracted from the Leibniz hyperpyramid \eqref{hiperpiramide_Leibniz} in the way 

\begin{equation}r^{(\nu)}_{N,n,m,l}=\frac{r^{(1)}_{N+4(\nu-1),n+\nu-1,m+\nu-1,l+\nu-1}}{{r^{(1)}_{4(\nu-1),\nu-1,\nu-1,\nu-1}}}\label{hiperpiramide_Leibniz_nu}\end{equation}
for any positive integer $\nu$, which satisfy a furtherly generalized Leibniz rule
\begin{equation}r^{(\nu)}_{N,n,m,l}+r^{(\nu)}_{N,n+1,m-1,l}+r^{(\nu)}_{N,n,m-1,l+1}+r^{(\nu)}_{N,n,m-1,l}=r^{(\nu)}_{N-1,n,m-1,l},\label{invariancia_escala_hiperpiramides}\end{equation}
which states that adding up certain four elements subpyramids of the $N-$th pyramid (analogous to the subpyramids of the Leibniz-like pyramid depicted in Fig. \ref{figura_tetraedro}) one gets a corresponding element of the $(N-1)-$th pyramid. Relation \eqref{invariancia_escala_hiperpiramides} is the $d=3$ instance of relation \eqref{scale_invariance_general}.

In order to get an explicit expression for coefficients \eqref{hiperpiramide_Leibniz_nu}, we shall first rewrite Leibniz-like triangle coefficients \eqref{triangulos_nu} in the form 
\begin{equation}r^{(\nu)}_{N,n}=\frac{\Gamma(2\nu)}{\Gamma^2(\nu)}\frac{\Gamma(N-n+\nu)\Gamma(n+\nu)}{\Gamma(N+2\nu)}\label{triangulo_bis},\end{equation}
and correspondingly, coefficients \eqref{tetraedro_beta} of the Leibniz-like pyramids in the form 
\begin{equation}r^{(\nu)}_{N,n,m}=\frac{\Gamma(3\nu)}{\Gamma^3(\nu)}\frac{\Gamma(N-n-m+\nu)\Gamma(n+\nu)\Gamma(m+\nu)}{\Gamma(N+3\nu)},\label{tetraedro_bis}\end{equation}
which naturally leads us to rewrite the coefficients  \eqref{hiperpiramide_Leibniz_nu} of the family of hyperpyramids as
\begin{equation}r^{(\nu)}_{N,n,m,l}=\frac{\Gamma(4\nu)}{\Gamma^4(\nu)}\frac{\Gamma(N-n-m-l+\nu)\Gamma(n+\nu)\Gamma(m+\nu)\Gamma(l+\nu)}{\Gamma(N+4\nu)}\end{equation}
which satisfy relation \eqref{invariancia_escala_hiperpiramides} for any $\nu>0$. 

Generalizing former results, the corresponding family of probability distributions 
\begin{equation}p^{(\nu)}_{N,n,m,l}=\binom{N}{n,m,l}r^{(\nu)}_{N,n,m,l}\end{equation}
associated to random variable $\vec X^{(\nu)}_{(3)}=\sum_{i=1}^N\vec X^{(\nu)}_i$,with $\langle\vec X^{(\nu)}_{(3)}\rangle=\left(\frac{N}{4}, \frac{N}{4}, \frac{N}{4}\right)$ for all $\nu$ and covariance matrix
\begin{equation}\Sigma^{(\nu)}_{(3)}=\frac{N(N+4\nu)}{16(1+4\nu)}\left(\begin{array}{ccc}
3&-1&-1\\
-1&3&-1\\
-1&-1&3\end{array}\right)\end{equation}
yields a tetranomial distribution, with $p_1=p_2=p_3=1-p_1-p_2-p_3=\frac{1}{4}$, in the Boltzmann limit: 
\begin{equation}p^{(\infty)}_{N,n,m,l}=\lim_{\nu\to\infty}p^{(\nu)}_{N,n,m,l}=\binom{N}{n,m,l}\frac{1}{4^N}\label{tetranomial_nu_infinito}\end{equation} 
where independent dices are recovered.

We are finally prepared to extend the above structure to higher dimensions. To describe a random experiment consisting of throwing a $(d+1)$-sided dice $N$ times, we need a $d-$dimensional random variable
$\vec X_{(d)}=(X_1,X_2,\dots,X_d)$, with 
\begin{equation}\vec X_{(d)}=\vec X_1+\vec X_2+\dots+\vec X_N\end{equation}
where $\vec X_j=(X_{1,j},X_{2,j},\dots,X_{d,j})$, $j=1,\dots,N$, is the  $d$-dimensional, $(d+1)$-valued random variable associated to the $j-$th throw, taking values $\vec e_1\equiv(1,0,\dots,0)$ for side labelled 1, $\vec e_2\equiv(0,1,\dots,0)$ for side labelled 2, $\dots$, $\vec e_d\equiv(0,\dots,1)$ for side labelled $d$, and $\vec 0\equiv(0,\dots,0)$ for side labelled $(d+1)$ ($d=1$ corresponds to binary variables, $d=2$ corresponds to ternary variables, and so on). Thus, variable $X_i=\sum_{j=1}^NX_{i,j}$, for $i=1,\dots,d$,  counts the number of appearances of side labelled $i$ out of $N$ throwings. 

Variable $\vec X_{(d)}$ follows a multinomial distribution
\begin{equation}P(X_1=n_1,X_2=n_2,\dots,X_d=n_d)=\binom{N}{n_1,n_2,\dots,n_d}p_1^{n_1}p_2^{n_2}\cdots p_d^{n_d}(1-p_1-\cdots-p_d)^{1-n_1-\cdots-n_d}\end{equation}    
where $p_i$ stands for the probability of obtaining side $i$,  $0\leqslant n_1+n_2+\cdots+n_d\leqslant N$, and the multinomial coefficients, $\binom{N}{n_1,n_2,\dots,n_d}=\frac{N!}{n_1!n_2!\cdots n_d!(N-n_1-\cdots-n_d)!}$, stand for the different ways to obtain the result $(n_1,n_2,\dots,n_d)$. Thus, there are only $\frac{(N+1)(N+2)\cdots(N+d)}{d!}$ different probability values, namely $r_{N,n_1,n_2,\dots,n_d}=p_1^{n_1}p_2^{n_2}\cdots p_d^{n_d}(1-p_1-\cdots-p_d)^{1-n_1-\cdots-n_d}$ among the $\Omega(N)=(d+1)^N$ events of the sample space. 

The $(d+1)$-dimensional Leibniz hyperpyramid is given by
\begin{equation}r^{(1)}_{N,n_1,n_2,\dots,n_d}=\frac{d!}{(N+1)(N+2)\dots(N+d)\binom{N}{n_1,n_2,\dots,n_d}}
\label{hiper_piramide_Leibniz_general}\end{equation}
from which the family of hyperpyramids
\begin{align}r^{(\nu)}_{N,n_1,n_2,\dots,n_d}&=\frac{r^{(1)}_{N+(d+1)(\nu-1),n_1+\nu-1,n_2+\nu-1,\dots,n_d+\nu-1}}{r^{(1)}_{(d+1)(\nu-1),\nu-1,\nu-1,\dots,\nu-1}}\nonumber\\&=\frac{\Gamma((d+1)\nu)}{\Gamma(\nu)^{d+1}}\frac{\Gamma(N-n_1-n_2-\cdots-n_d+\nu)\Gamma(n_1+\nu)\Gamma(n_2+\nu)\cdots\Gamma(n_d+\nu)}{\Gamma(N+(d+1)\nu)}\label{hiper_piramide_Leibniz_general_nu}\end{align}
with associated probabilities 
\begin{equation}p^{(\nu)}_{N,n_1,n_2,\dots,n_d}=\binom{N}{n_1,n_2,\dots,n_d}r^{(\nu)}_{N,n_1,n_2,\dots,n_d}\end{equation}
for variable $\vec X^{(\nu)}_{(d)}=\sum_{i=1}^N\vec X^{(\nu)}_i$ is obtained, with $\langle\vec X^{(\nu)}_{(d)}\rangle=\left(\frac{N}{d+1},\dots,\frac{N}{d+1}\right)$ for all $\nu$, and covariance matrix
\begin{equation}\Sigma^{(\nu)}_{(d)}=\frac{N(N+(d+1)\nu)}{(d+1)^2(1+(d+1)\nu)}\left(\begin{array}{cccc}
d&-1&\cdots&-1\\
-1&d&\ddots&\vdots\\
\vdots&\ddots&\ddots&-1\\
-1&\cdots&-1&d\end{array}\right)\end{equation}
(concerning the component variables $X^{(\nu)}_{i,j}$, $i=1,\dots,d$. $j=1,\dots,N$, one has $\langle X^{(\nu)}_{i,j}\rangle=\frac{1}{d+1}$, $\sigma^2_{X^{(\nu)}_{i,j}}=\frac{d}{(d+1)^2}$, $\sigma^2_{X^{(\nu)}_{i,j}X^{(\nu)}_{i,k}}=r^{(\nu)}_{2,0,\dots,0}-\frac{1}{(d+1)^2}=\frac{d}{(d+1)^2(1+(d+1)\nu)}$, for $j\neq k$, $\sigma^2_{X^{(\nu)}_{i,j}X^{(\nu)}_{l,j}}=-\frac{1}{(d+1)^2}$, for $i\neq l$, and $\sigma^2_{X^{(\nu)}_{i,j}X^{(\nu)}_{k,l}}=r^{(\nu)}_{2,1,0,\dots,0}-\frac{1}{(d+1)^2}=-\frac{1}{(d+1)^2(1+(d+1)\nu)}$, for $i\neq k$ and $j\neq l$.) 

The generalized scale invariance rule, followed by coefficients \eqref{hiper_piramide_Leibniz_general_nu} for $d\geqslant 2$, is given in  Eq.~\eqref{scale_invariance_general}.

\section{Conclusions}
\label{conclusions}

We have generalized to an arbitrary dimension a one dimensional discrete probabilistic model first proposed in \cite{RodriguezSchwammleTsallis} which, for one dimension, yields $q-$Gaussians in the thermodynamic limit. These functions, which play a central role in nonextensive statistical mechacics, appear in a plethora of applications to natural, social and artificial systems. 

Though we have obtained two-dimensional distributions which contain one-dimensional conditional  distributions yielding one-dimensional $q-$Gaussians in the thermodynamic limit, our calculations seem to indicate that the model does not yield, for generic $q$, bidimensional $q-$Gaussians as limiting probability distributions for $N\to\infty$. In addition, making use of the corresponding generalization of Eq.~\eqref{condicionales_relacion}, our model contains one-dimensional $q-$Gaussians as conditional probability distributions for any starting dimension $d$ (corresponding in turn to sums of $(d+1)-$ary variables). 

We are thus lead to the conclusion that the case of binary variables is special. Indeed, it appears to be the only one yielding $q-$Gaussians in the thermodynamic limit in a simple manner. For more complex random variables, an adaptation seems necessary in what concerns the boundaries that emerge.  More precisely, a  
difficulty exists to match the domain where the random variables of our model take their values (e.g., a triangle or a tetrahedron for dimensions 2 or 3 respectively) with the support of the $q-$Gaussians, which is the interior of an ellipse or of an ellipsoid in dimensions 2 or 3 respectively, for the case $q<1$; or the whole euclidian space for $q>1$ (in which case, as shown in Ref.~\cite{HanelThurnerTsallis} for dimension 1, a complicated change of variables is needed in order to transform the bounded support of the discret model into the whole real axis). This difficulty becomes apparent when trying to use a sort of generalization of the Laplace-de Finetti theorem \cite{deFinetti}, which allows for a representation of exchangeable stochastic processes, and has been successfully applied to the case of binary random variables \cite{HanelThurnerTsallis}.

Though the formulation of a simple probabilistic model yielding multidimensional $q-$Gaussians in the thermodynamic limit still remains an  open question, we have introduced a very rich scale-invariant multidimensional model which further support the hypothesis that scale invariance is possibly a necessary ---though definitely not sufficient--- condition for $q-$independence, which in turn guarantees the appearence of $q-$Gaussian attractors.

\begin{acknowledgments}
The authors thank V. Schwammle and P. Tempesta for helpful comments, as well as partial financial support by CNPq and FAPERJ (Brazilian Agencies) and DGU-MEC (Spahish Ministry of Education) through Projects PHB2007-0095-PC and MODELICO.  

\end{acknowledgments}

\section*{Appendix}

In order to prove relation \eqref{condicionales_relacion}, we will substitute Eq.~\eqref{condicionales}, with $p^{(\nu)}_{N,n,m}$ given in \eqref{distribucion_generalizada}, in the l.h.s. and Eq.~\eqref{probabilidad_nu} in the r.h.s. of \eqref{condicionales_relacion}, to obtain
\begin{equation}P^{(\nu)}_{N,n|m}=\frac{\displaystyle\binom{N}{n,m}r^{(\nu)}_{N,n,m}}{\displaystyle\sum_{n=0}^{N-m}\binom{N}{n,m}r^{(\nu)}_{N,n,m}}=\frac{\displaystyle\binom{N-m}{n}r^{(\nu)}_{N,n,m}}{\displaystyle\sum_{n=0}^{N-m}\binom{N-m}{n}r^{(\nu)}_{N,n,m}}=\binom{N-m}{n}r^{(\nu)}_{N-m,n}\end{equation}
where we have made use of the relation $\binom{N}{n,m}=\binom{N}{m}\binom{N-m}{n}$. Thus, Eq.~\eqref{condicionales_relacion} is fulfilled whenever 
\begin{equation}\frac{r^{(\nu)}_{N,n,m}}{\displaystyle\sum_{n=0}^{N-m}\binom{N-m}{n}r^{(\nu)}_{N,n,m}}=r^{(\nu)}_{N-m,n}\label{triangulos_renormalizados}\end{equation}
which establishes a relationship between properly rescaled elements of the pyramid and the elements of the triangle with the same value of $\nu$.

Introducing now in \eqref{triangulos_renormalizados} expressions \eqref{triangulo_bis} and \eqref{tetraedro_bis} for the coefficients of the triangles and pyramids respectively, and eliminating terms which cancel out, one gets

\begin{equation}\frac{1}{\displaystyle\sum_{n=0}^{N-m}\binom{N-m}{n}\Gamma(N-n-m+\nu)}=\frac{\Gamma(2\nu)}{\Gamma(\nu)^2\Gamma(N+2\nu)}
\end{equation}
which, making again use of Eq.~\eqref{triangulo_bis}, is trivially fulfilled
\begin{align}1&=\frac{\Gamma(2\nu)}{\Gamma(\nu)^2\Gamma(N+2\nu)}\sum_{n=0}^{N-m}\binom{N-m}{n}\Gamma(N-n-m+\nu)\Gamma(n+\nu)\nonumber\\&=\sum_{n=0}^{N-m}\binom{N-m}{n}r^{(\nu)}_{N-m,n}=\sum_{n=0}^{N-m}p^{(\nu)}_{N-m,n}\end{align}

\bibliography{basename of .bib file}

\newpage

\end{document}